\documentclass[12pt]{iopart}  

\usepackage{color,amssymb,graphicx,hyperref,iopams,ulem}
\definecolor{darkblue}{rgb}{0.0,0.0,0.3}
\hypersetup{colorlinks,breaklinks,
            linkcolor=darkblue,urlcolor=darkblue,
            anchorcolor=darkblue,citecolor=darkblue}
\begin{document}
\newcommand{\vv}{\boldsymbol{v}}
\newcommand{\w}{\boldsymbol{w}}

\bibliographystyle{unsrt}

\title[Drug Binding and Dynamic Instability]{Modeling the Effects of Drug
Binding on the Dynamic Instability of Microtubules}
\author{Peter Hinow$^1$, Vahid Rezania$^2$, Manu Lopus$^3$,  Mary Ann Jordan$^3$
and Jack A.~Tuszy\'nski$^4$}

\address{$^1$Department of Mathematical Sciences, University of
Wisconsin -- Milwaukee, P.O.~Box 413, Milwaukee, WI 53201, USA} 
\ead{hinow@uwm.edu}
\address{$^2$Department of Physical Sciences, Grant MacEwan University,
Edmonton AB, T5J 4S2, Canada}
\address{$^3$Department of Molecular,
Cellular, and Developmental Biology and the Neuroscience Research Institute,
University of California, Santa Barbara, CA 93106, USA}
\address{$^4$Cross Cancer Institute and
Department of Physics, University of Alberta, Edmonton AB, T6G 2J1, Canada}
\begin{abstract}
\begin{sloppypar}
We propose a stochastic model that accounts for the growth, catastrophe and
rescue processes of steady state microtubules assembled from MAP-free tubulin
in the possible presence of a microtubule associated drug. As an example for the
latter, we both experimentally and theoretically study the perturbation of
microtubule dynamic instability by  \mbox{S-methyl-D-DM1}, a synthetic
derivative of the microtubule-targeted agent maytansine and a potential
anticancer agent. Our model predicts that among drugs that act
locally at the microtubule
tip, primary inhibition of the loss of GDP tubulin results in stronger damping
of microtubule dynamics than inhibition of GTP tubulin addition. On the other
hand, drugs whose action occurs in the interior of the microtubule need to be
present in much higher concentrations to have visible effects.
\end{sloppypar}

\end{abstract}
\submitto{\PB}
\maketitle

\section{Introduction}
Microtubules are hollow and flexible cylindrical polymers of the protein tubulin
that form a major component of the cytoskeleton of eukaryotic cells. They play a
central role in maintenance of structural stability of the cell, intracellular
vesicle transport and chromosome separation during mitosis. The polymerization
of  tubulin into microtubules and the subsequent catastrophic
depolymerization have been studied extensively both experimentally and
theoretically, see \cite{Jackson,Mandelkow89,Walker88,BLT99,BLT99b,Mitchison,
FHL94,FHL96,JWF97,Rezania_BPJ,Lopus09,vanBuren2002,vanBuren2005,PRE} for just a
few examples. The prevailing model to explain \textit{dynamic instability}, the
lateral cap model, is that a cap of GTP tubulin at the growing tip is required
for stability of the polymer and that a loss of this GTP cap results in dramatic
shortening of microtubules \cite{Bayley2}.

In the paper \cite{PRE} we proposed a partial differential equation
model inspired by dynamics of size-structured populations. The variables of
that continuous model are length distributions of microtubules and the amounts
of free tubulin. The major reactions are the polymerization of free GTP
tubulin, the hydrolysis of assembled GTP tubulin to GDP tubulin, the decay and
rescue of microtubules without a GTP cap and the recycling of free GDP tubulin
to GTP tubulin. The model conserves the total amount of tubulin in all its
forms. In addition, we allowed for the nucleation of fresh microtubules at
certain specified (short) lengths. With a small number of parameters that have
clear biochemical interpretations, we were able to reproduce commonly observed
experimental behaviors, such as oscillations in the amount of tubulin assembled
into microtubules. Obviously, the continuous model is not expected to
reproduce the lengths of individual microtubules. To this end, in this paper we
present a stochastic discrete model of microtubule dynamic instability and
compare its predictions to observations of microtubule lengths from \textit{in
vitro} experiments that also include the presence of a dynamic instability
suppressing drug. 

Stochastic discrete models of biopolymer dynamics have been investigated, among
others, in \cite{BLT99,BLT99b,Antal,Gregoretti,Matzavinos,vanBuren2002,
vanBuren2005, Brun,Mishra}. Bolterauer \textit{et al.}~\cite{BLT99,BLT99b}
introduced a stochastic model to study the dynamics of free microtubules using
the master equation approach. They showed that in the continuum limit the
microtubule length distribution follows a bell-shaped curve. Mishra \textit{et
al.}~\cite{Mishra} applied a similar technique to elaborate the effect of
catastrophe-suppressing drugs on the dynamic instability of microtubules. They
assumed that drug molecules bind rapidly to free tubulin in the solution. Then,
the drug-tubulin complexes bind to the growing tips of the microtubules and
reduce the catastrophe frequency by stabilizing the microtubule caps. They also
assumed that the drug-tubulin complex has a lower attachment rate than drug-free
tubulin. As a result, one would expect to observe shorter drug-treated
microtubules in the steady state. While they found a qualitative reduction in
catastrophe frequency, surprisingly, they found that the drug-treated
microtubules have the same length distribution as free microtubules. 

Here we present a stochastic model that represents the same set of reactions as
our continuous model in \cite{PRE}, except for the nucleation of fresh
microtubules which is an unnecessary complication for the case studied here. 
Similar to other stochastic models \cite{Antal,Gregoretti,
vanBuren2002,vanBuren2005,Brun,Mishra}, our underlying model that describes
microtubule dynamics without drugs is based on an effective two-state model,
switching between a growing and shortening state. As in \cite{PRE} and in
contrast to other works such as \cite{Antal, Mishra}, we couple the growth
velocity of microtubules to the amount of free GTP tubulin and the model is
constructed so as to preserve the total amount of tubulin in all its forms.
Following \cite{FHL96}, we include hydrolysis events of two
kinds, namely \textit{scalar hydrolysis} (conversion of bound GTP tubulin to GDP
tubulin with equal probability) and \textit{vectorial hydrolysis} (the
probability is enhanced by a GDP tubulin neighbor). Furthermore, in contrast
to previous models that simulate one microtubule at a time, we simulate
several microtubules at the same time. This enables us to capture a more
realistic situation similar to one in experiments.  Last but not least, a
great advantage of our model is that it can be implemented
straightforwardly with the help of the Gillespie Algorithm \cite{Gillespie}, an
exact simulation method for stochastic chemical reactions.
Our goal is to explain individual length observations of microtubules growing
without microtubule associated proteins (MAPs) \textit{in vitro}
or length constraints (as in \cite{Gregoretti}). Moreover, we
investigate the effects of tubulin-binding drugs on microtubule dynamic
instability. These drugs belong roughly to one of two classes, namely those that
bind to assembled microtubules and those that bind to free tubulin
\cite{Jordan07}. This difference in behaviors may be due to conformational
changes of the $\alpha/\beta$-tubulin heterodimer \cite{Bennett} (the tubulin
\textit{unit} from now on) upon incorporation into the microtubule lattice that
expose or hide the binding site for the drug molecule. Apart from these
different binding modes, the effects on the microtubule reactions can also
differ \cite{Correia08}. Some drugs mainly slow down microtubule formation while
others mainly prevent microtubule decay. This is not a strict dichotomy in that
some drugs can have multiple actions, depending on their concentration. For
example, vinblastine inhibits microtubule formation at high drug concentrations,
and inhibits microtubule decay at low concentrations \cite{Jordan07}. Moreover,
there are multiple action mechanisms of drugs, for example either by preventing
addition of GTP tubulin,loss of GDP tubulin from an exposed tip or by preventing
the hydrolysis of GTP tubulin incorporated in the microtubule. In any case,
within a living cell exposed to anti-mitotic agents, mitosis cannot be completed
and the cell dies. This property of tubulin-binding drugs leads to many
successful anti-cancer chemotherapeutics such as paclitaxel, vincristine,
vinblastine to name but a few. Maytansine and its derivatives  are known to
suppress microtubule dynamics \textit{in vitro} and in cells \cite{Oroudjev}. As
an example, we focus on the potential anticancer agent S-methyl-D-DM1, a
synthetic derivative of the microtubule-binding agent maytansine. Antibody-DM1
conjugates are currently under clinical trials with promising results
\cite{Lopus11}. In our model, drugs act by decelerating (or
accelerating) binding or release reactions by a certain factor. Thus the
strength of the action can be quantified and linked to  
the binding energy through the standard Arrhenius relation. 

\section{The stochastic model}
We consider a linear model of the microtubule and disregard the fact that it
actually consists of 13-17 protofilaments arranged in a helical lattice.
Tubulin can be added to the microtubule in form of small oligomers of varying
sizes \cite{Mozziconacci08}.

\begin{figure}[th]
\begin{center}
\includegraphics[width=80mm]{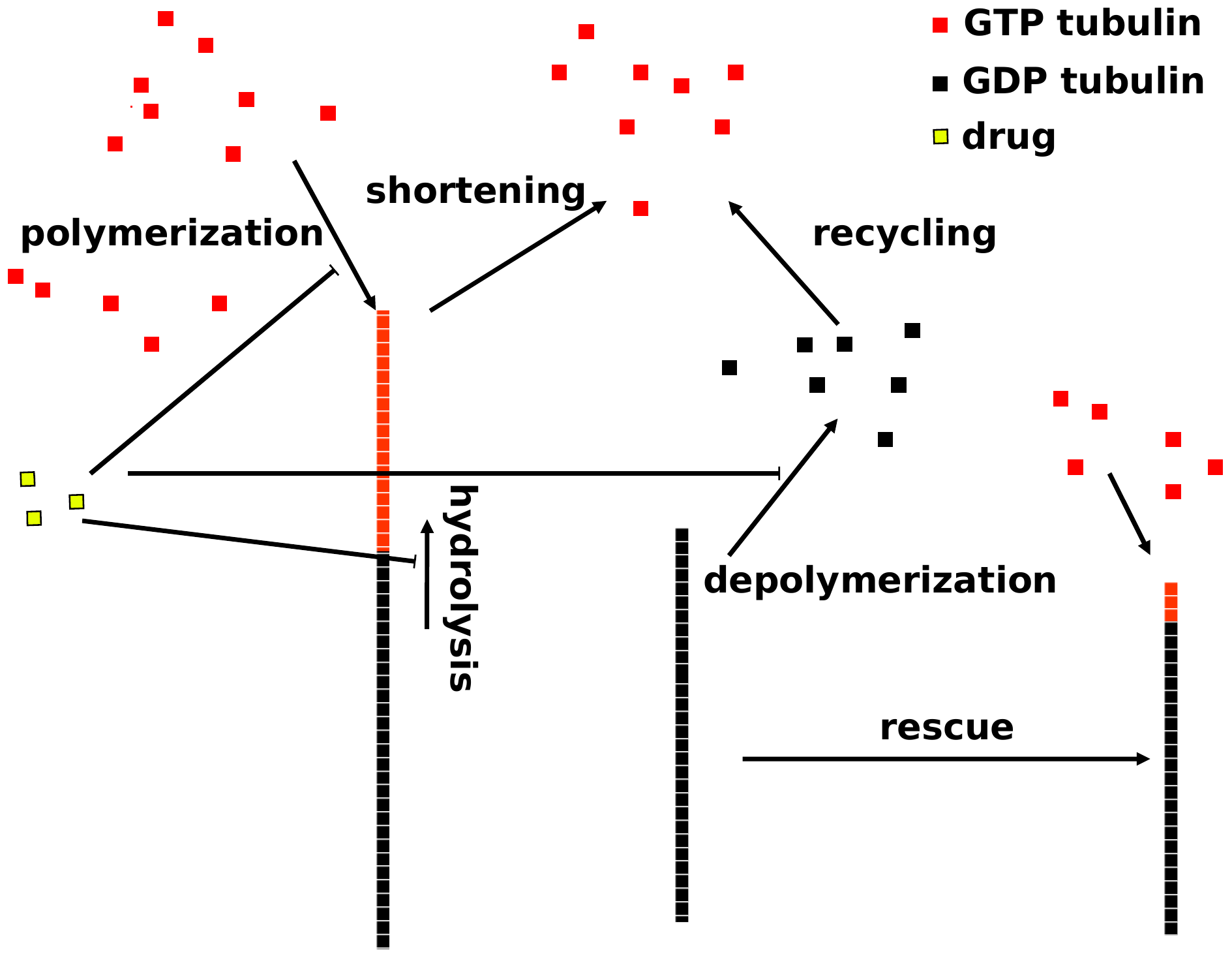}
\caption{The reactions and possible drug interactions
implemented in our stochastic model. A drug may also promote polymerization and
depolymerization.}\label{reaction_scheme}
\end{center}
\end{figure}

\begin{figure}[th]
\begin{center}
\includegraphics[width=65mm, height=27mm]{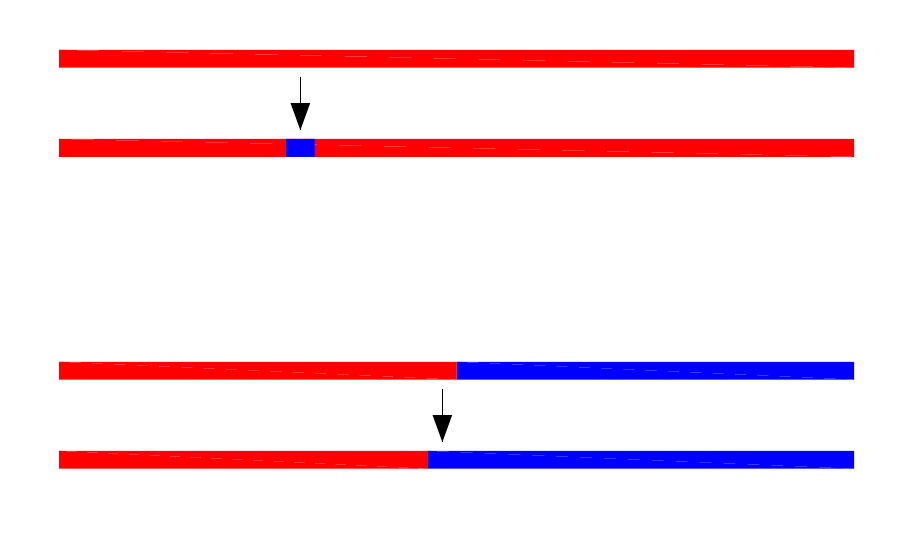}
\includegraphics[width=60mm]{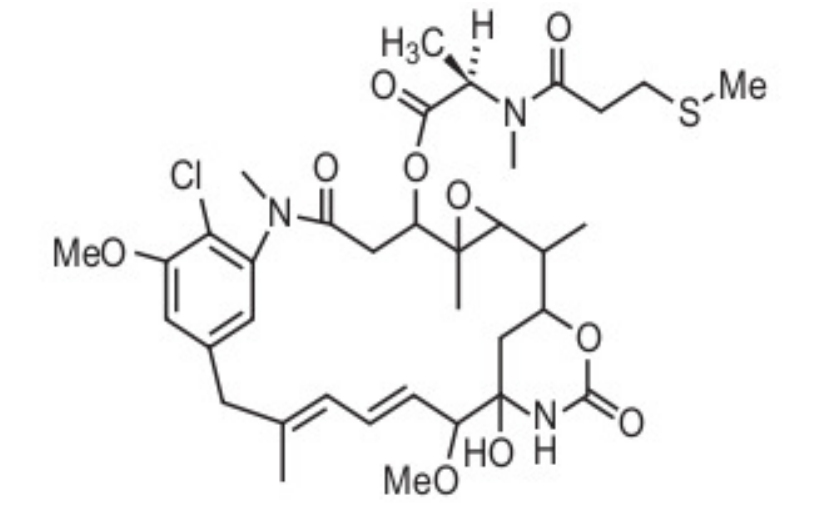}
\caption{(Left panel) Schematic depiction of the two hydrolysis mechanisms. The
scalar hydrolysis reaction (p.~4 (\ref{scalar}), top) picks a random bound GTP
tubulin and changes it into a bound GDP tubulin. The vectorial hydrolysis
reaction (p.~5 (\ref{vectorial}), bottom) occurs at a boundary between a GDP
zone and a GTP zone. (Right panel) Structural formula of the maytansine analog
S-methyl-D-DM1.}\label{figure1}
\end{center}
\end{figure}

Let $m\ge 1$ be the number of dynamic microtubules. The state of each
microtubule is represented as a word $\vv^k=(\dots,v^k_2,v^k_1,v^k_0)$, $k=
1,\dots,m$ on the binary alphabet $\{0,1\}$ where the letter $1$ stands for
a position occupied by a GTP tubulin unit and $0$ stands for a position
occupied by a GDP tubulin unit. The length of the microtubule $\vv^k$ is denoted
by $|\vv^k|$. The number of GTP tubulin units within  $\vv^k$ is denoted by
$I(\vv^k)$. The ``tip'' of the microtubule is the letter $v^k_0$ and this is the
only position where growth or shrinkage can occur. Consecutive strings of 0s and
1s are called GDP zones and GTP zones, respectively. The number of boundaries
between such zones is denoted by $B(\vv^k)$. The numbers of free GTP tubulin and
GDP tubulin are denoted by $N^T$ and $N^D$, respectively. These numbers can be
converted to and from concentrations, if the volume in which the reactions take
place is given. The following reactions occur (see also Figure
\ref{reaction_scheme}).
\begin{enumerate}
\item\label{growth} Growth by attachment of GTP tubulin(s) 
\begin{equation*}
\vv^k \mapsto [\vv^k\, \underbrace{1\,\dots \,1}_{l} ],\quad N^T\mapsto N^T-l,
\end{equation*}
at rate  (derived from mass action kinetics) 
\begin{equation}\label{normal_binding}
\lambda N^T(v^k_0+p(1-v^k_0)).
\end{equation}
 Notice that every microtubule has its own growth reaction so
that there are $m$ of them. 
The number of added GTP tubulin units $l$ can be set to a fixed value (say, 1)
or drawn from a Poisson distribution with parameter $L$. The dimensionless
parameter $p\ge 0$ is the propensity of a rescue event when a GDP tubulin  unit
at the tip of the microtubule is exposed. In the simplest case, $p=1$,
attachment of a new GTP tubulin is independent of the tip status. 
\item\label{loss_GDP} Loss of a GDP tubulin  (when the
tip is in the state 0) 
\begin{equation*}
\vv^k \mapsto (\dots,v^k_2,v^k_1),\quad  N^D\mapsto N^D+1,
\end{equation*}
at rate $\mu_{GDP}(1-v^k_0)$. Again there are $m$ such
shrinking reactions.
\item\label{loss_GTP} Loss of a GTP tubulin (when the
tip is in the state 1) 
\begin{equation*}
\vv^k \mapsto (\dots,v^k_2,v^k_1),\quad  N^T\mapsto N^T+1,
\end{equation*}
at rate $\mu_{GTP}v^k_0$. 
\item\label{scalar} 
\begin{sloppypar}
Hydrolytic conversion of a bound GTP tubulin
$\vv^{k^*}\mapsto\widetilde{\vv}^{k^*}$ at rate $\delta_{sc} \sum_{k=1}^m
I(\vv^k)$. The index $k^*$ is chosen uniformly in the set $\{1,\dots,m\}$ and a
random position of $\vv^{k^*}$ that is occupied by $1$ is changed to $0$ (this
hydrolysis mechanism is called \textit{scalar} hydrolysis in \cite{FHL96}). 
\item\label{vectorial} Hydrolytic conversion of a bound GTP tubulin
$\vv^{k^*}\mapsto\widetilde{\vv}^{k^*}$ at rate $\delta_{vec} 
\sum_{k=1}^mB(\vv^k)$. Again, the index $k^*$ is chosen uniformly in the set
$\{1,\dots,m\}$ and the word $\widetilde{\vv}^{k^*}$ is created by selecting
randomly a position of $\vv^{k^*}$ where a $1$ neighbors a $0$ and changing that
$1$ to $0$ (this hydrolysis mechanism is called \textit{vectorial} hydrolysis in
\cite{FHL96}, see Figure \ref{figure1}, left panel). 
\end{sloppypar}
\item\label{recycling} Recycling of free GDP tubulin to GTP tubulin
\begin{equation*}
N^D \mapsto N^D-1,\quad N^T\mapsto N^T+1,
\end{equation*}
at rate $\kappa N^D$. It is assumed that a sufficient amount of chemical energy
in the form of free GTP is always present.
\end{enumerate}

This scheme can be simplified by setting some parameter values to zero. For
example, one may disregard the possibility of a bound GTP tubulin to be lost
again ($\mu_{GTP}=0$, cf.~\cite{FHL96}), although other authors argue that this
may take place in up to $90\%$ of all binding events \cite{Howard09}. The
addition size $l$ can be selected to be constant 1. The hydrolysis reaction
(\ref{scalar}) picks any bound GTP tubulin and changes it to a GDP tubulin,
thereby creating islands of GTP tubulin within the length of the microtubule.
That this is possible and important for the rescue process was recently shown by
Dimitrov \textit{et al.}~\cite{Dimitrov}. Both hydrolysis mechanisms in concert
provide an indirect coupling of the hydrolysis reaction to the addition of new
GTP tubulin units \cite{FHL96}. 

Tubulin-binding drugs can bind to tubulin in one of two states, whether it is
free or bound within a microtubule. Here, we consider drugs that suppress
microtubule dynamic instability by specifically binding to microtubules. For
every microtubule encoded by a word $\vv$, we introduce a second word
$\w=(\dots,w_2,w_1,w_0)$ over the alphabet $\{0,1\}$ (the drug state), of equal
length as $\vv$. Here $w_i=1$, if the tubulin unit at position $v_i$ is occupied
by a drug molecule and $w_i=0$ otherwise. There are binding events of drug
molecules to unoccupied sites and release of drug molecules from the
microtubule. Let $E(\w)$ be the number of available sites for drug binding
and let $F(\w)$ be the number of drug occupied sites. The latter is always the
sum of the entries $1$ in $\w$ while the former may be only a subset of entries
$0$ in $\w$. We have the association and dissociation events 
\begin{enumerate}
\item[(vii)]\label{bind_drugB} Binding of drug to tubulin units within the
microtubule
\begin{equation*}
\w^{k^*} \mapsto \widetilde{\w}^{k^*},\quad  D\mapsto D-1,
\end{equation*}
at rate $\rho D\sum_{k=1}^mE(\w^k)$. The new word $\widetilde{\w}^{k^*}$ is
obtained by selecting randomly one letter $0$ among the sites available for
binding and changing it to $1$. This set may be the set of all entries $0$ or
the entries $0$ that are within a certain distance from the tips or the
unoccupied tips alone.
\item[(viii)]\label{release_drugB} Release of drug from the occupied sites of
the microtubule 
\begin{equation*}
\w^{k^*} \mapsto \widetilde{\w}^{k^*},\quad D\mapsto D+1,
\end{equation*}
at rate $\sigma\sum_{k=1}^mF(\w^k)$. The new word $\widetilde{\w}^{k^*}$ is
obtained by changing one randomly selected letter $1$ in any of the drug words
to $0$. 
\end{enumerate}
The reactions (\ref{growth})--(\ref{recycling}) have the same outcomes as
far as changes in numerical quantities are concerned, however the rates of
reactions (\ref{growth}), (\ref{loss_GDP}) and (\ref{loss_GTP}) have a more
complicated dependence upon the status of the microtubule tip. Since the tip can
now have four different states, the attachment process (\ref{growth}) to
microtubule $k$ occurs at rate  
\begin{equation}\label{drug_growth}
\lambda N^T\left(\big(v^k_0+p(1-v^k_0)\big)(1-w^k_0)
+r\big(v^k_0+p(1-v^k_0)\big)w^k_0\right), 
\end{equation}
where the dimensionless non-negative constant $r$ modulates the attachment
propensity compared to the drug-free tip, see Equation (\ref{normal_binding}).
Small values of $r$ would mean that attachment of new GTP tubulin units is
hindered by drug molecules bound to the tip. On the other hand, values $r>1$
would increase microtubule polymerization. The shrinking reactions
(\ref{loss_GDP}) and (\ref{loss_GTP}) occur at rates 
\begin{equation}\label{drug_shrink}
\mu_{GDP}(1-v^k_0)\big((1-w^k_0)+qw^k_0\big), \quad \textrm{and}
\quad \mu_{GTP}v^k_0\big((1-w^k_0)+qw^k_0\big), 
\end{equation}
where a small value of $q\ge0$ implies a high level of protection afforded by a
drug molecule bound to the tip. If a drug bound tubulin can fall off a
microtubule (i.e.~if $q>0$), then the drug-tubulin compound is assumed to
dissociate immediately. We refer to Figure
\ref{reaction_scheme} for the possible interactions of the
drug with the microtubules (as implemented in this paper).

There are also possible drug actions beyond the microtubule tips. As was first
discovered by Lin and Hamel \cite{Lin1981}, a drug can also inhibit the
hydrolysis of bound GTP tubulin. This can be implemented by ``splitting'' the
scalar and vectorial hydrolysis reactions (\ref{scalar}) and
(\ref{vectorial}). More precisely, let $I_0(\vv^k)$ and $I_1(\vv^k)$ be the
number of GTP tubulin units within the microtubule word $\vv^k$ that are
unoccupied respectively occupied by a drug molecule. Then the scalar hydrolysis
reactions (iv${}^0$) and (iv${}^1$) occur at rates $\delta_{sc} \sum_{k=1}^m
I_0(\vv^k)$ (as before) and $s\delta_{sc} \sum_{k=1}^m I_1(\vv^k)$ (with $0\le
s\le1$). A similar splitting is used for the vectorial hydrolysis reaction
(\ref{vectorial}). 

\section{Materials and Methods}
\begin{sloppypar}
Tubulin ($15\,\mu M$), phosphocellulose purified, MAP-free, was assembled on the
ends of sea urchin (\textit{Strongylocentrotus purpuratus}) axoneme fragments at
$30^\circ C$ in $87\, mmol/L$ Pipes, $36\, mmol/L$ Mes, $1.4 \,mmol/L$
MgCl${}_2$, $1\, mmol/L$ EGTA, pH 6.8 (PMME buffer) containing $2 \,mmol/L$ GTP
for 30 $min$ to achieve steady state. We used a $100 \,nmol/L$ concentration of
S-methyl-D-DM1
($N^{2'}$-deacetyl-$N^{2'}$-(3-thiomethyl-1-oxopropyl)-D-maytansine
\cite{Widdison}, see Figure \ref{figure1}, right panel), which had no
considerable effect on microtubule polymer mass, to analyze their individual
effects on dynamic instability. Time-lapse images of microtubule plus ends were
obtained at $30^\circ C$ by video-enhanced differential interference
contrast microscopy at a spatial resolution of $0.3\,\mu m$ using an Olympus
IX71 inverted microscope with a 100 $\times$ oil immersion objective (NA = 1.4).
The end of an axoneme that possesses more, faster growing, and longer
microtubules than  the other end was designated as the plus end as described
previously \cite{Lopus10a, Lopus10b}. Microtubule dynamics were recorded
for 40 $min$ at $30^\circ C$, capturing $\sim 10\, min$ long videos for each
area under observation. The microtubules were tracked approximately every 3 $s$ using RTMII software, and
the life-history data were obtained using IgorPro software (MediaCybernetics,
Bethesda, MD) \cite{Yenjerla}. 
\end{sloppypar}

We have programmed the reactions (\ref{growth})--(viii) 
using the Gillespie Algorithm \cite{Gillespie} (in \textsc{Java}, available from
the corresponding author upon request). This algorithm simulates the chemical
reactions as collisions of particles in real time and its parameters are the
actual reaction rates, not probabilities. If the empty word is reached, then a
new microtubule is created (a word containing a single GTP tubulin unit). The
gain in length due to addition of a single tubulin unit is taken to be
$\ell=8\,nm/13 = 0.6\,nm$; see e.g.~\cite[Equation (2)]{FHL96}.

\section{Results}
\begin{figure}[th]
\begin{center}
\includegraphics[width=60mm]{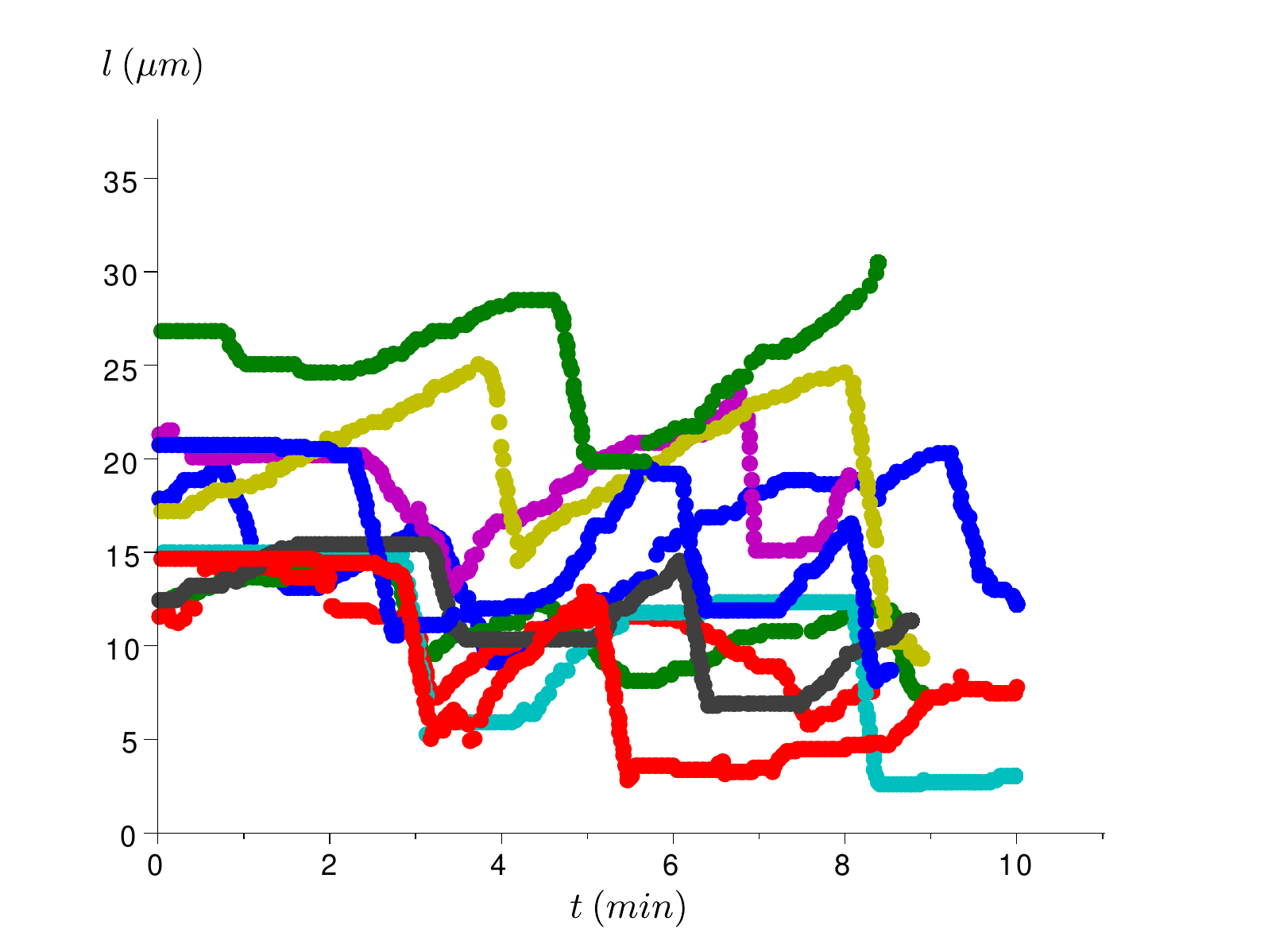}
\includegraphics[width=60mm]{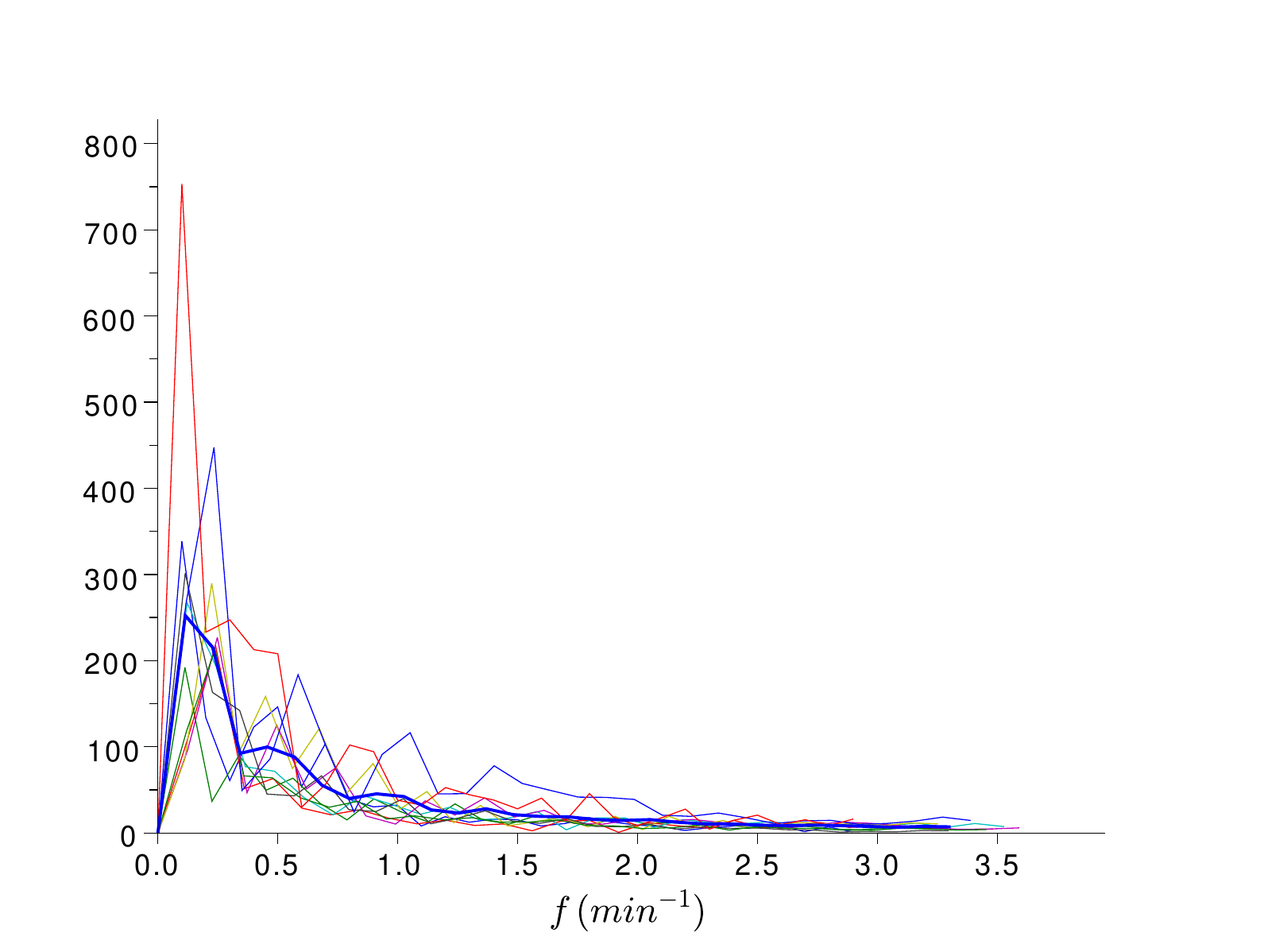}
\caption{(Left panel) Length time series of 10 microtubules in absence of drug.
Notice that life histories from several experiments are plotted in the same
diagram. (Right panel) Absolute Fourier spectra of the control experimental data
that were normalized to mean length zero. The thick blue line is the average of
the 10 individual spectra. }\label{control_data}
\end{center}
\end{figure}

Results of the \textit{in vitro} experiments are shown in Figures
\ref{control_data} and \ref{drug_data}. During periods of growth, the untreated
microtubules grow at a velocity of $\approx 3\,\mu m\,min^{-1}$
while during periods of shrinkage, microtubules shrink at $\approx 20\,\mu
m\,min^{-1}$. While they are infrequent, we attribute
occasional periods of stagnation to the spatial resolution of the microscope of
approximately 400 tubulin dimer units.

\begin{figure}[th]
\begin{center}
\includegraphics[width=60mm]{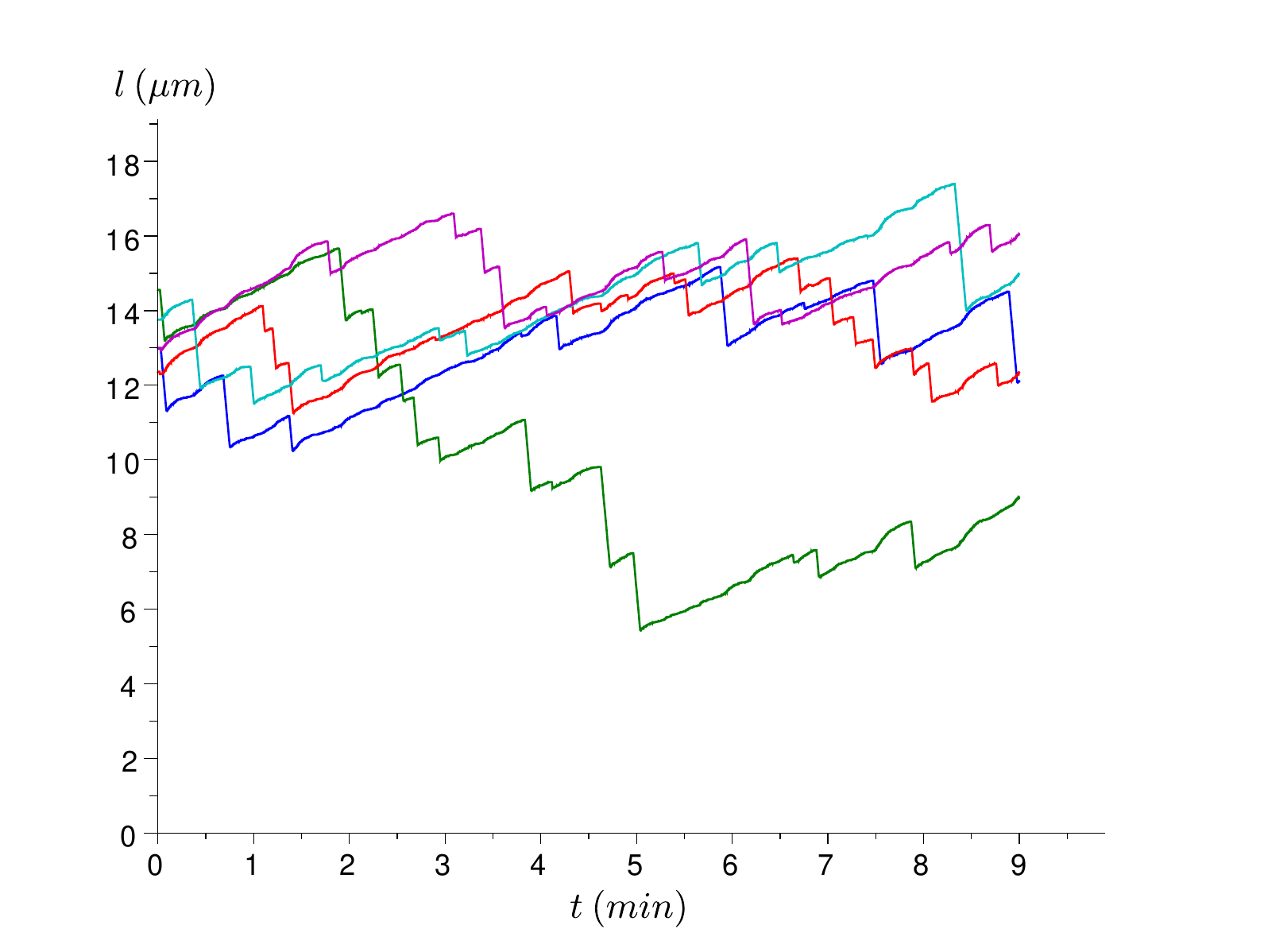}
\includegraphics[width=60mm]{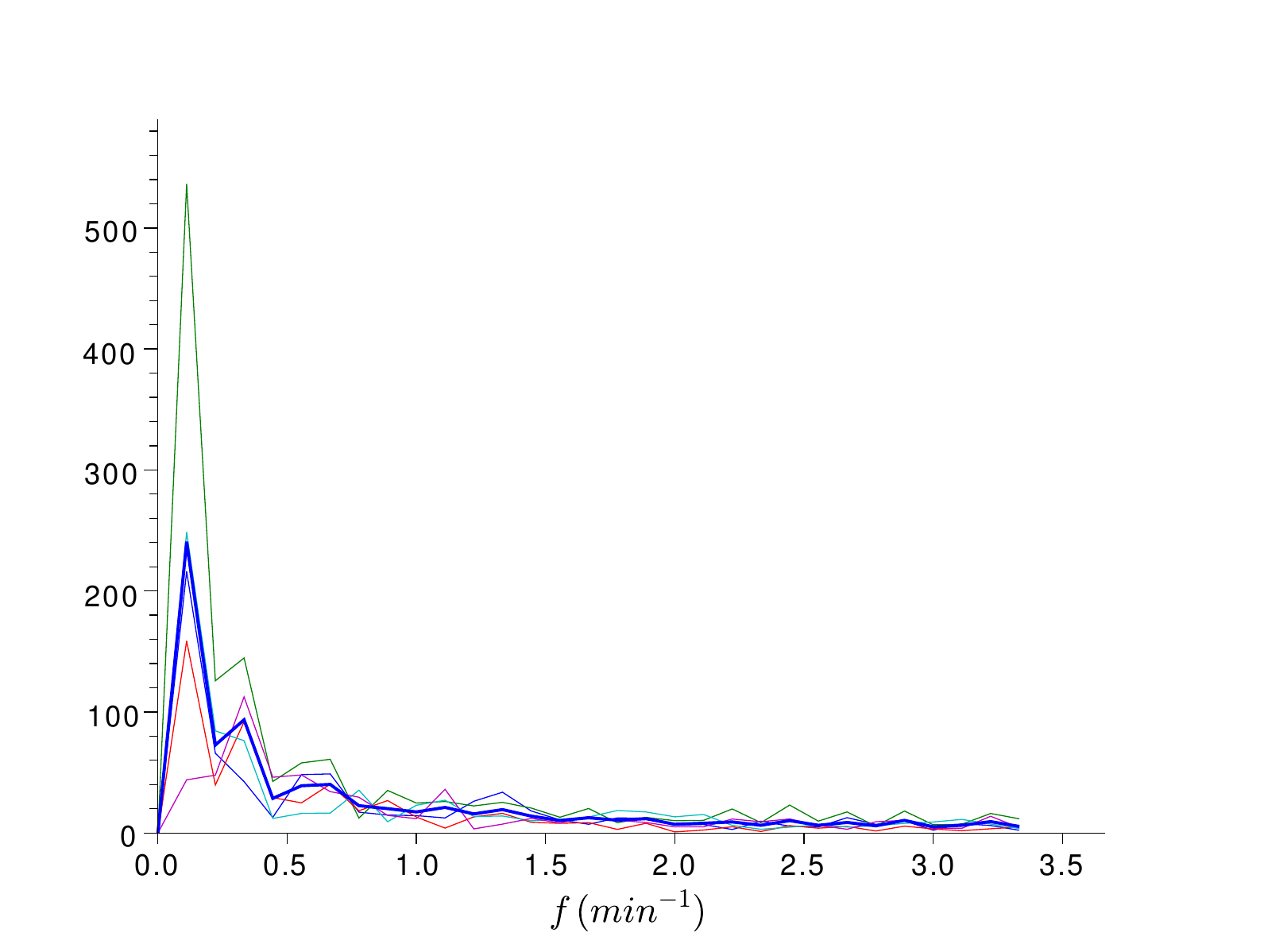}
\caption{(Left panel) Simulation of $m=5$ microtubules starting from random
initial states, with a total of $\approx 10^5$ tubulin units. The parameter
values  are as in Table \ref{Tab1}. GTP tubulin is added in the form
of oligomers whose length is Poisson distributed with with mean $L=6$. The
resulting average growth velocity during periods of growth is
approximately $2\,\mu m\,min^{-1}$ while the resulting
shrinking velocity is approximately
$20\,\mu m\,min^{-1}$. (Right panel) Absolute Fourier spectra of the simulation
data, normalized to mean length zero. The thick blue line is the average of the
5 individual spectra.}\label{control_simulation_random}
\end{center}
\end{figure}

\begin{figure}[th]
\begin{center}
\includegraphics[width=60mm]{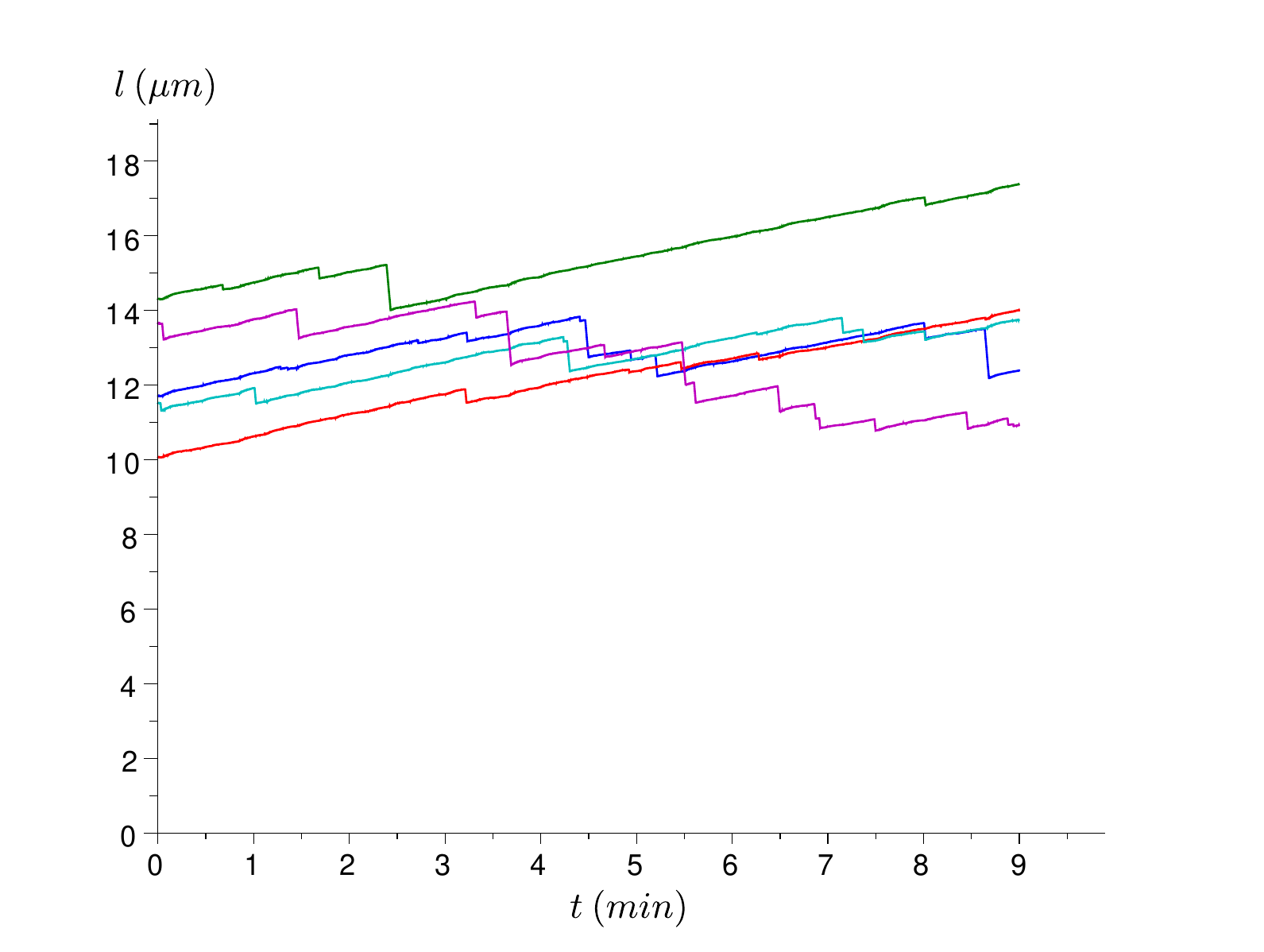}
\includegraphics[width=60mm]{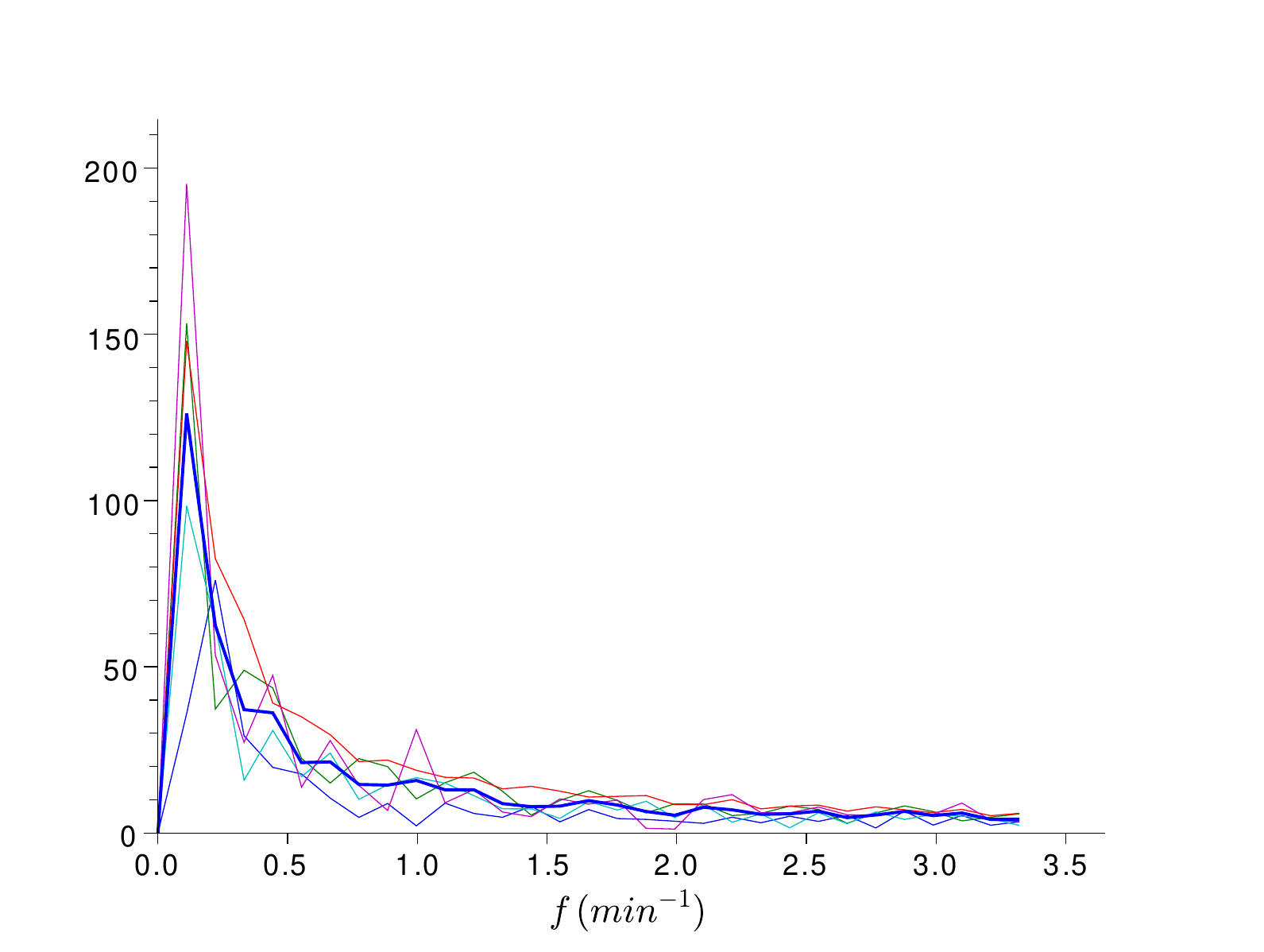}
\caption{(Left panel) Simulation of $m=5$ microtubules with parameter
values as in Table \ref{Tab1} and Figure \ref{control_simulation_random} except
that GTP tubulin is added in units of fixed length $L=1$.  (Right panel)
Absolute Fourier spectra of the simulation data, normalized to mean length zero
and their average.}\label{control_simulation_fixed}
\end{center}
\end{figure}

\begin{figure}[ht]
\begin{center}
\includegraphics[width=60mm]{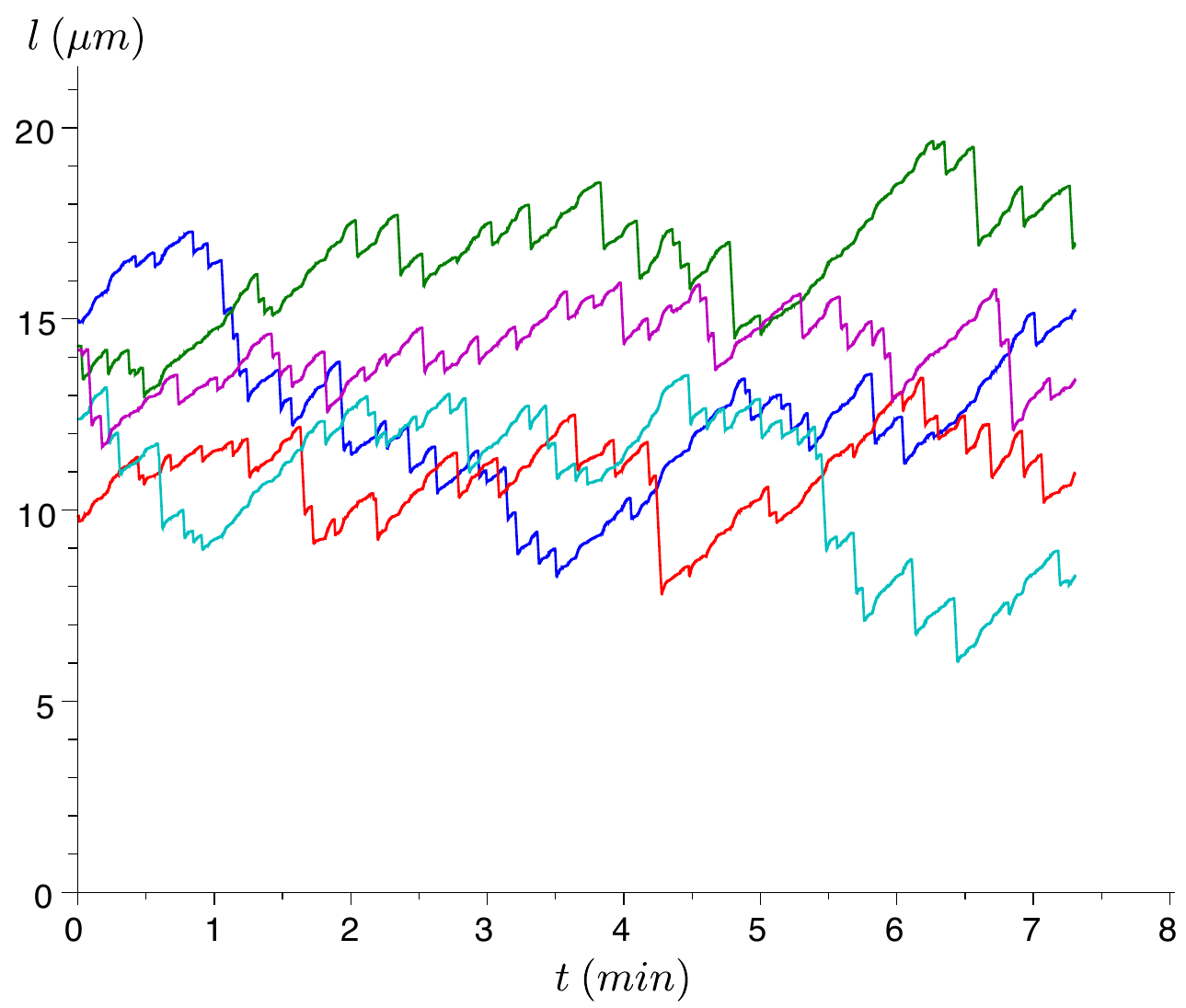}
\includegraphics[width=60mm]{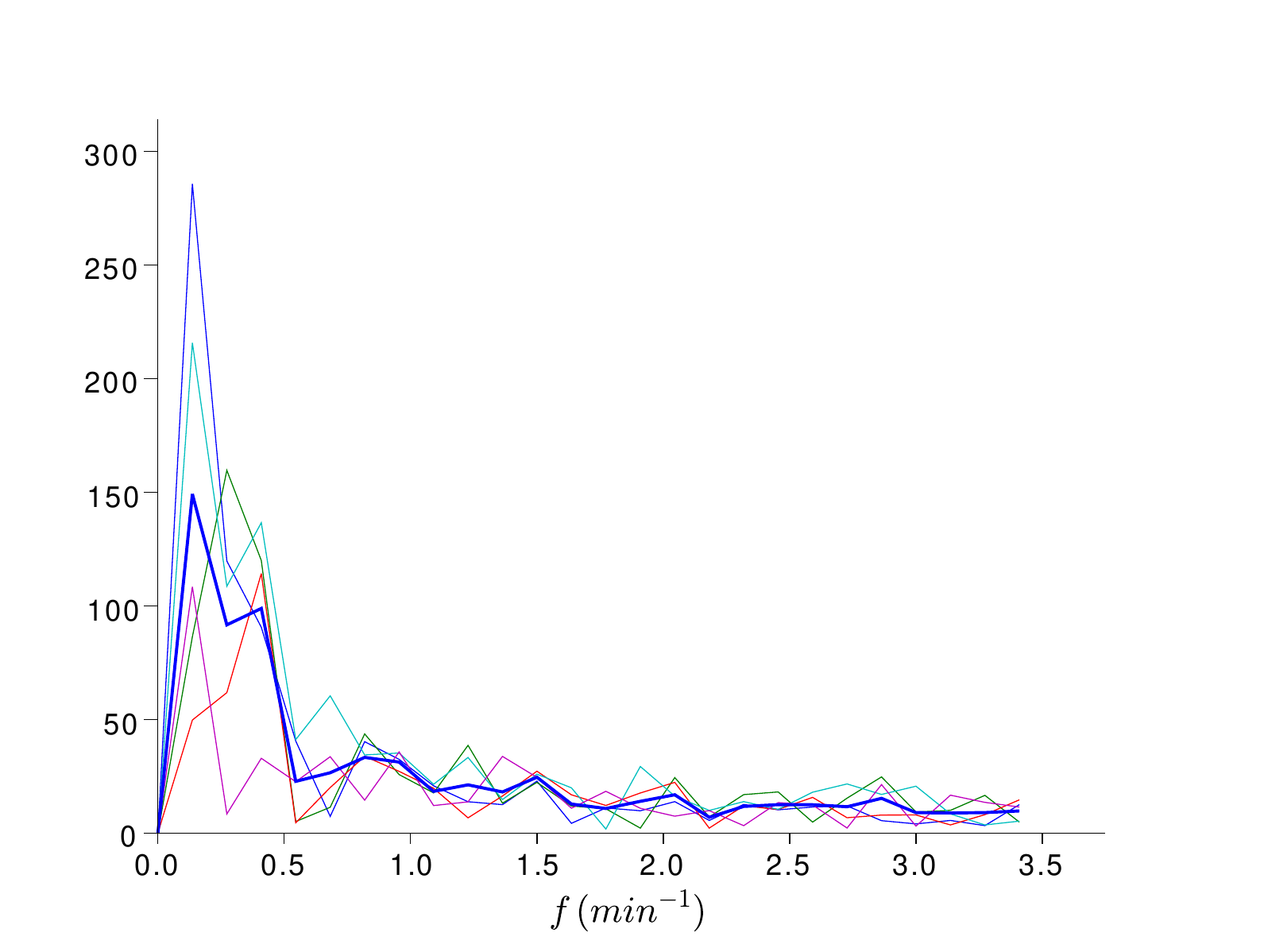}
\caption{(Left panel) Simulation of $m=5$ microtubules with parameter values
$\lambda=1.0 \,(\ell s)^{-1}$, $\mu_{GDP} = 2000\,s^{-1}$,
$\delta_{sc}=\delta_{vec}=3.0 \,(\ell s)^{-1}$ and  $\kappa=0.5\, s^{-1}$, all
of which are larger than those in Table \ref{Tab1}. (Right panel) The
corresponding absolute Fourier spectra show a visible shift towards higher
frequencies. }\label{simulation_faster}
\end{center}
\end{figure}

\begin{figure}[ht]
\begin{center}
\includegraphics[width=60mm]{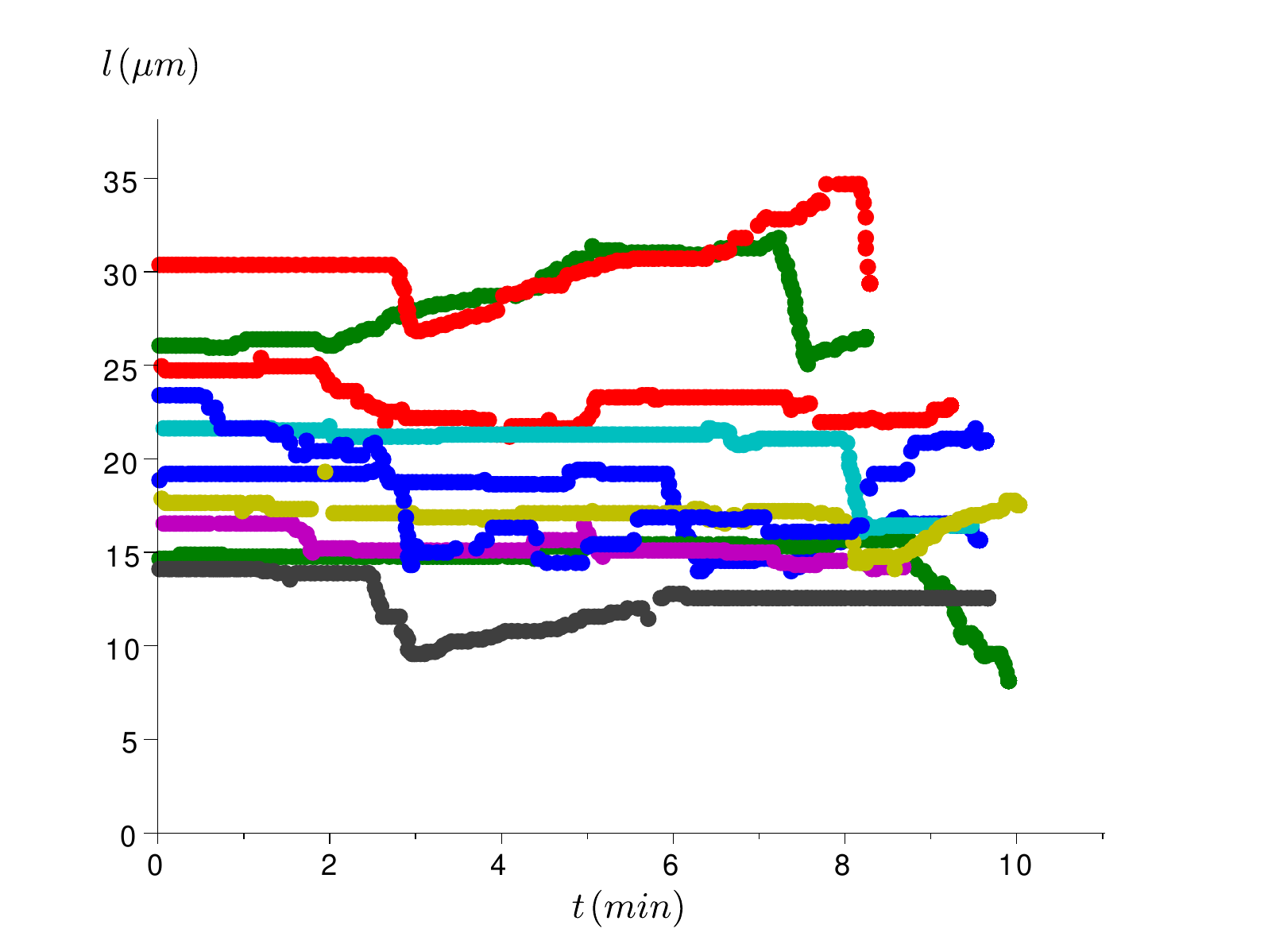}
\includegraphics[width=60mm]{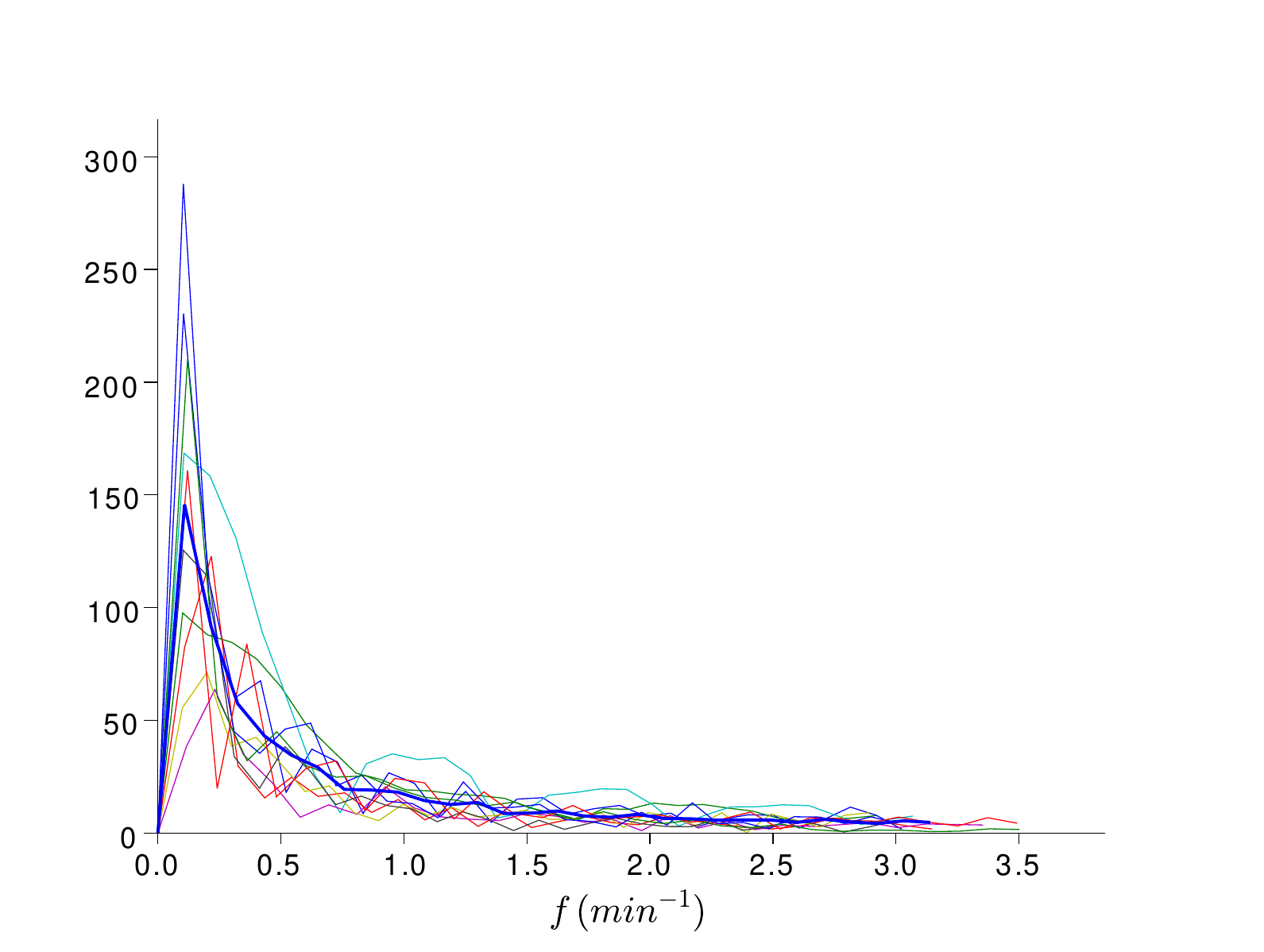}
\caption{(Left panel) Length time series of 10 microtubules in presence
of the drug S-methyl-D-DM1. (Right panel) The corresponding absolute Fourier
spectra, normalized to mean length zero. The thick blue line is the average of
the 10 individual spectra. }\label{drug_data}
\end{center}
\end{figure}

The left panels of Figures \ref{control_simulation_random},
\ref{control_simulation_fixed} and \ref{simulation_faster} show possible
simulations of the control scenario in the absence of drugs. The choice of
appropriate parameter values for the simulations is a difficult problem, since
different values have been reported in different literature sources and some
parameters have only been estimated on the basis of the Arrhenius Equation (see
\cite[Table 1]{PRE} for some ranges). While there is no unique choice of
parameter values, we observe a good agreement of the simulated and observed
growth and shrinking velocities, for the choice of parameter values in Table
\ref{Tab1}, see Figure \ref{control_simulation_random}. To quantify the
agreement between the simulations and the experimental data we used the absolute
Fourier spectra \cite{Odde1996} of the length time series. We first re-sampled
both the experimental data and the numerical simulations on equispaced time
grids at approximately $0.4\,Hz$. In order to make different simulations
comparable, we subtracted the mean from each length time series so that the
resulting normalized lengths have mean zero. If $l_n$, $n=0,\dots,N-1$ are the
re-sampled normalized lengths, then the discrete Fourier transform is given by
the absolute values of
\begin{equation*}
 L_k = \sum_{n=0}^{N-1} l_n \exp\left(-\frac{2\pi ikn}{N}\right), \quad k
=0,\dots,N-1.
\end{equation*}
This is conveniently done with the help of the fast Fourier transform routine
\texttt{fft} in \textsc{scilab}\footnote{Open source software; available
at \texttt{www.scilab.org}.}. Results are shown in the right panels of Figures
\ref{control_data} and \ref{control_simulation_random}, showing a good agreement
among experimental and simulated data of the location and the height of the peak
of the averaged spectra. In Figure \ref{simulation_faster}, a
sensitivity check is run with significantly larger parameter values $\lambda =
1.0\,(\ell s)^{-1},\, \mu_{GDP} = 2000\, s^{-1},\, \delta_{sc} = \delta_{vec} =
3.0\,(\ell s)^{-1}$ and \mbox{$\kappa = 0.5\, s^{-1}$}, see Table \ref{Tab1} for
comparison. As a result, higher oscillation frequencies would be expected in
the Fourier spectra due to higher rate of polymerization, depolymerization,
hydrolysis and recycling events. This is clearly shown in the right panel of
Figure  \ref{simulation_faster}, as the spectrum visibly shifts towards higher
frequencies.  

\begin{figure}[ht]
\begin{center}
\includegraphics[width=60mm]{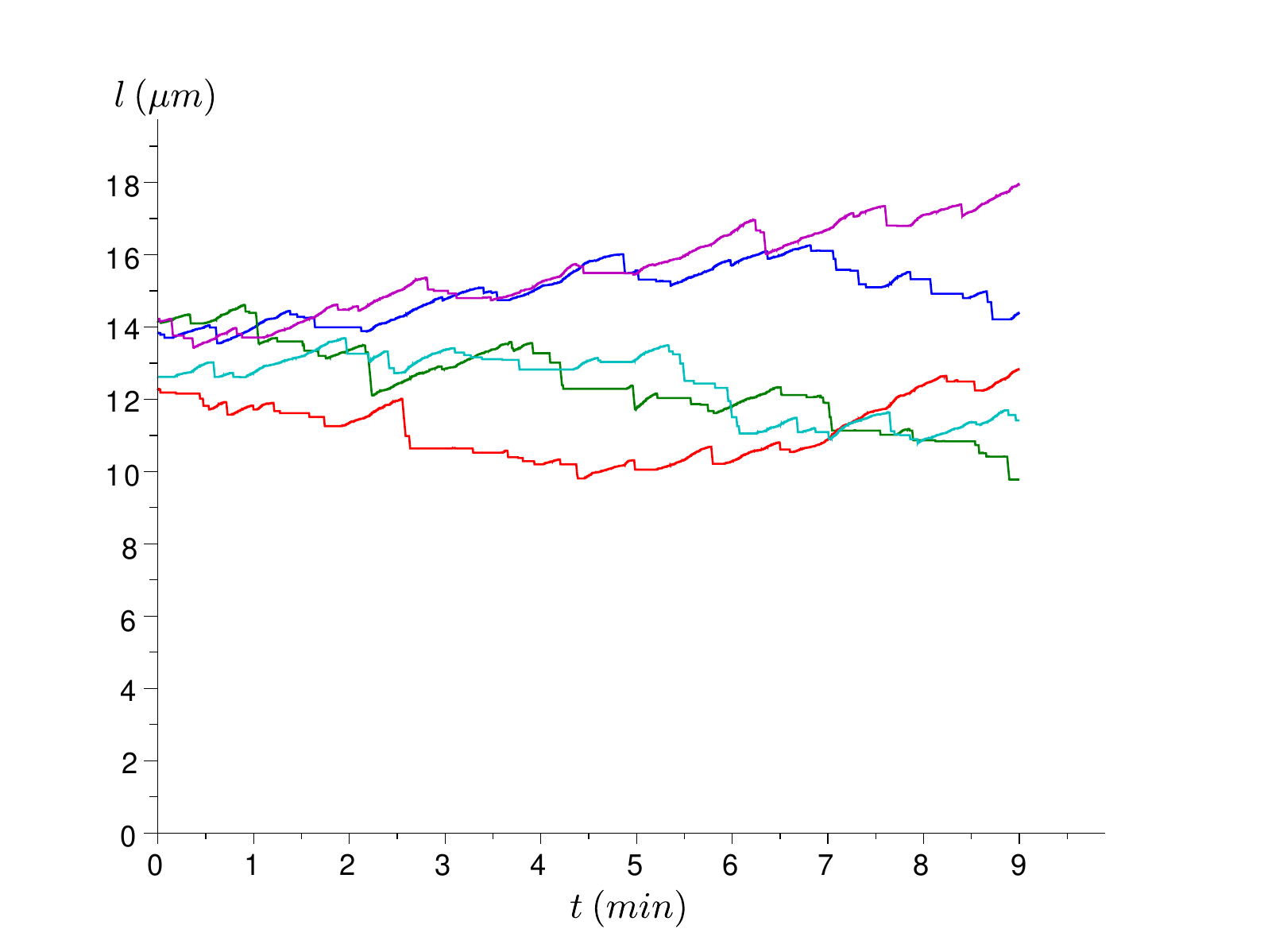}
\includegraphics[width=60mm]{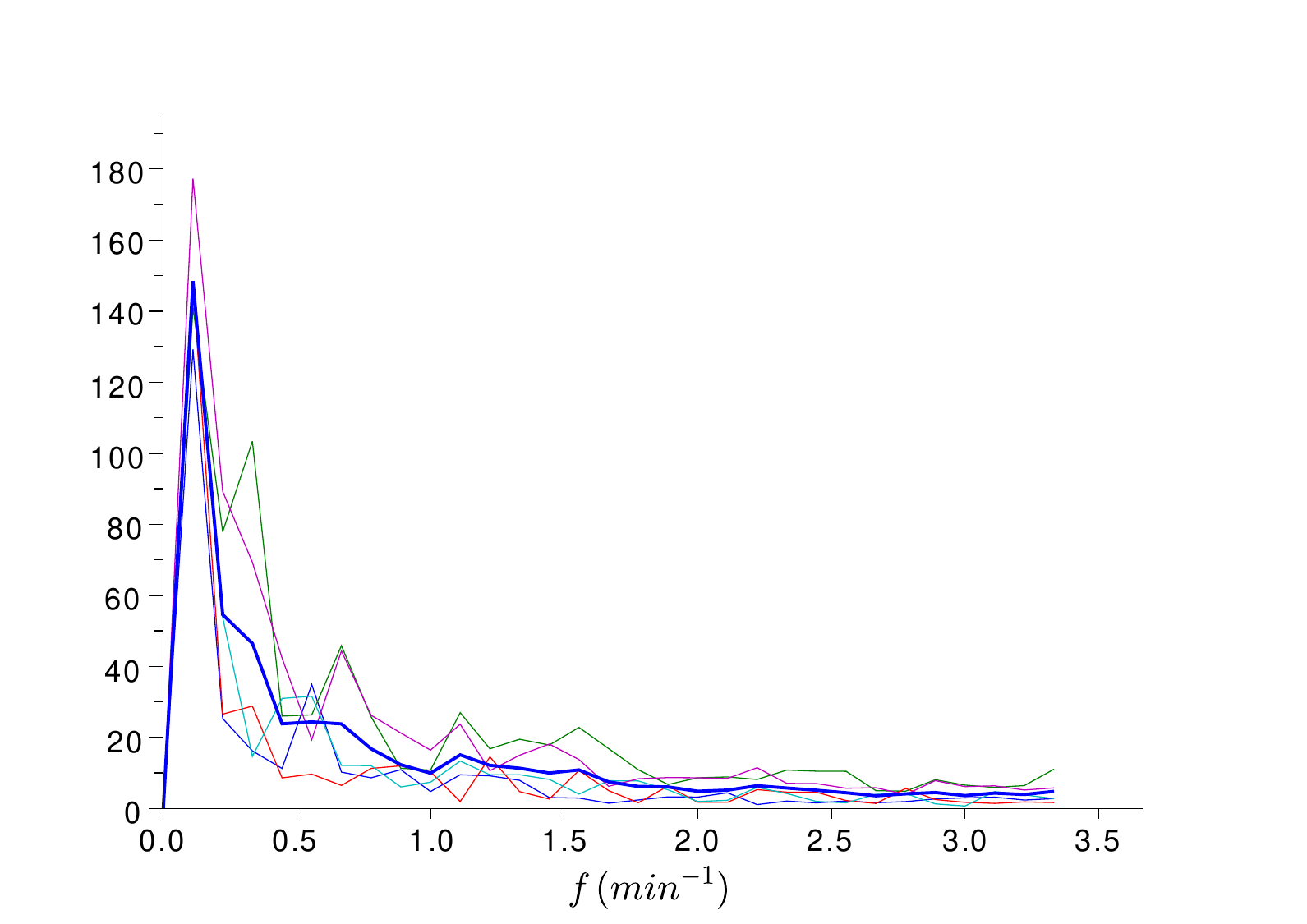}
\caption{(Left panel) Simulation of $m=5$ microtubules in the presence of $400$
drug molecules, with a total of $\approx 10^5$ tubulin units. The parameter
values  are  as in Figure \ref{control_simulation_random}, in addition $r=0.01$,
$q=0$ and $s=1$. Here the drug molecules bind to any open site
with equal probability. (Right panel) The corresponding absolute Fourier
spectra. This simulation suggests a
possible action mechanism for the drug S-methyl-D-DM1.}\label{drug_simulation}
\end{center}
\end{figure}

\begin{figure}[ht]
\begin{center}
\includegraphics[width=80mm]{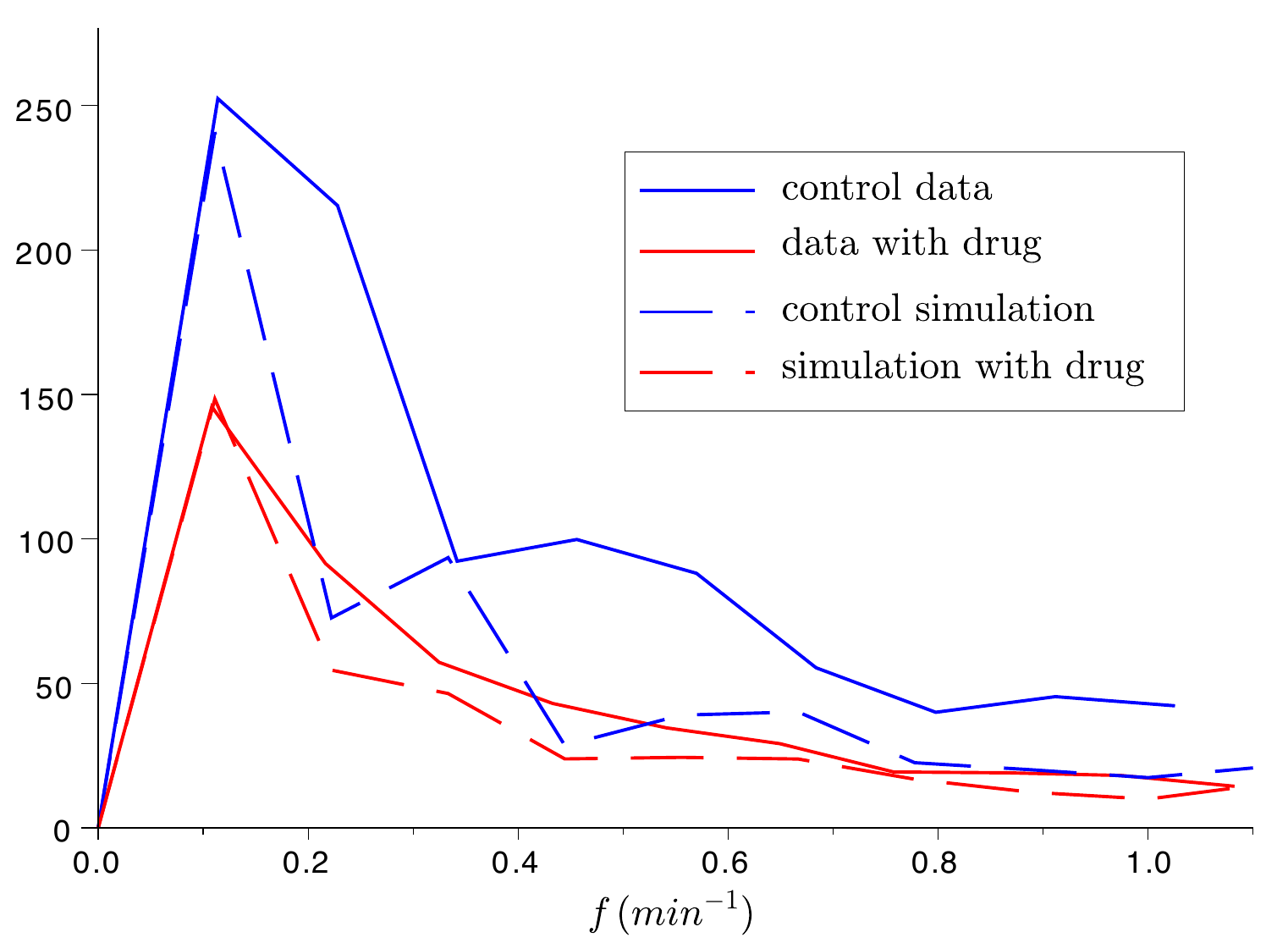}
\caption{Detail of the average absolute Fourier
spectra from Figures \ref{control_data}, \ref{control_simulation_random},
\ref{drug_data} and \ref{drug_simulation}. Within the resolution of the
frequency grid $\approx 0.11\,min^{-1}$, there is no discernible shift of
the position of the peaks.}\label{all_ffts}
\end{center}
\end{figure}

The presence of the drug clearly decelerates the
dynamic activity of the microtubules as can be seen from Figure \ref{drug_data}.
This is also visible in the reduced peak heights of the Fourier spectra,
while the position of the peak, i.e.~the main frequency of the oscillations
remains unchanged, at least not discernible on the $0.4\,Hz$
frequency grid, see also Figure \ref{all_ffts}. With the same parameter values
as in the control in Figure \ref{control_simulation_random}, we simulated the
presence of a microtubule-binding drug that suppresses strongly the addition of
new GTP tubulin ($r=0.01$), completely inhibits the loss of tubulin at the tip
($q=0$) and does not affect the hydrolysis of bound GTP tubulin ($s=1$). The
open drug binding sites are all entries $0$ in the drug words. The results are
shown in Figure \ref{drug_simulation}. There are 400 drug molecules present in
the simulation (compared to $\approx 10^5$ tubulin units). This amount of
roughly 100 drug molecules per microtubule matches approximately the 
concentration of S-methyl-D-DM1 in the experiments.

\begin{figure}[ht]
\begin{center}
\includegraphics[width=60mm]{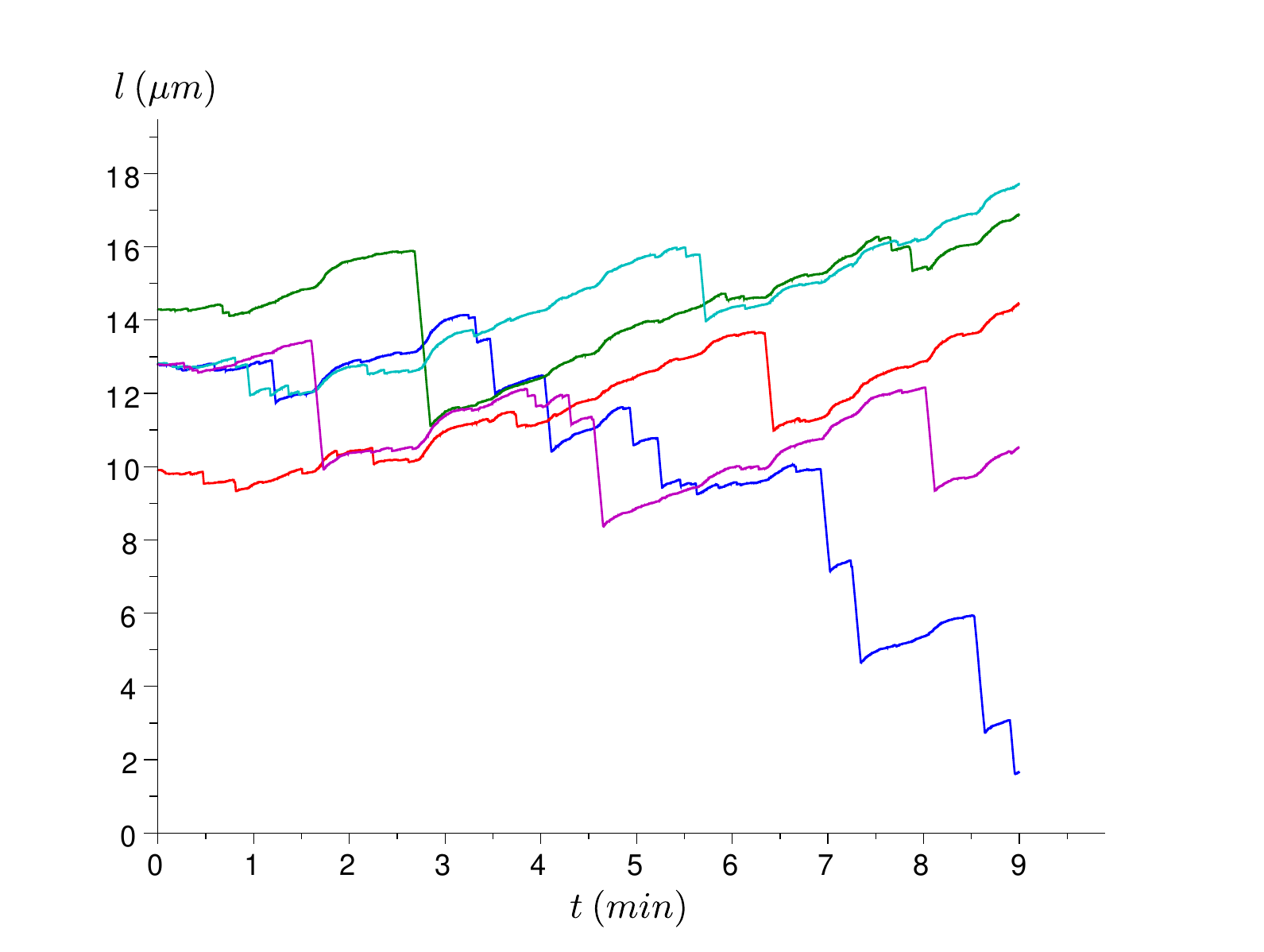}
\includegraphics[width=60mm]{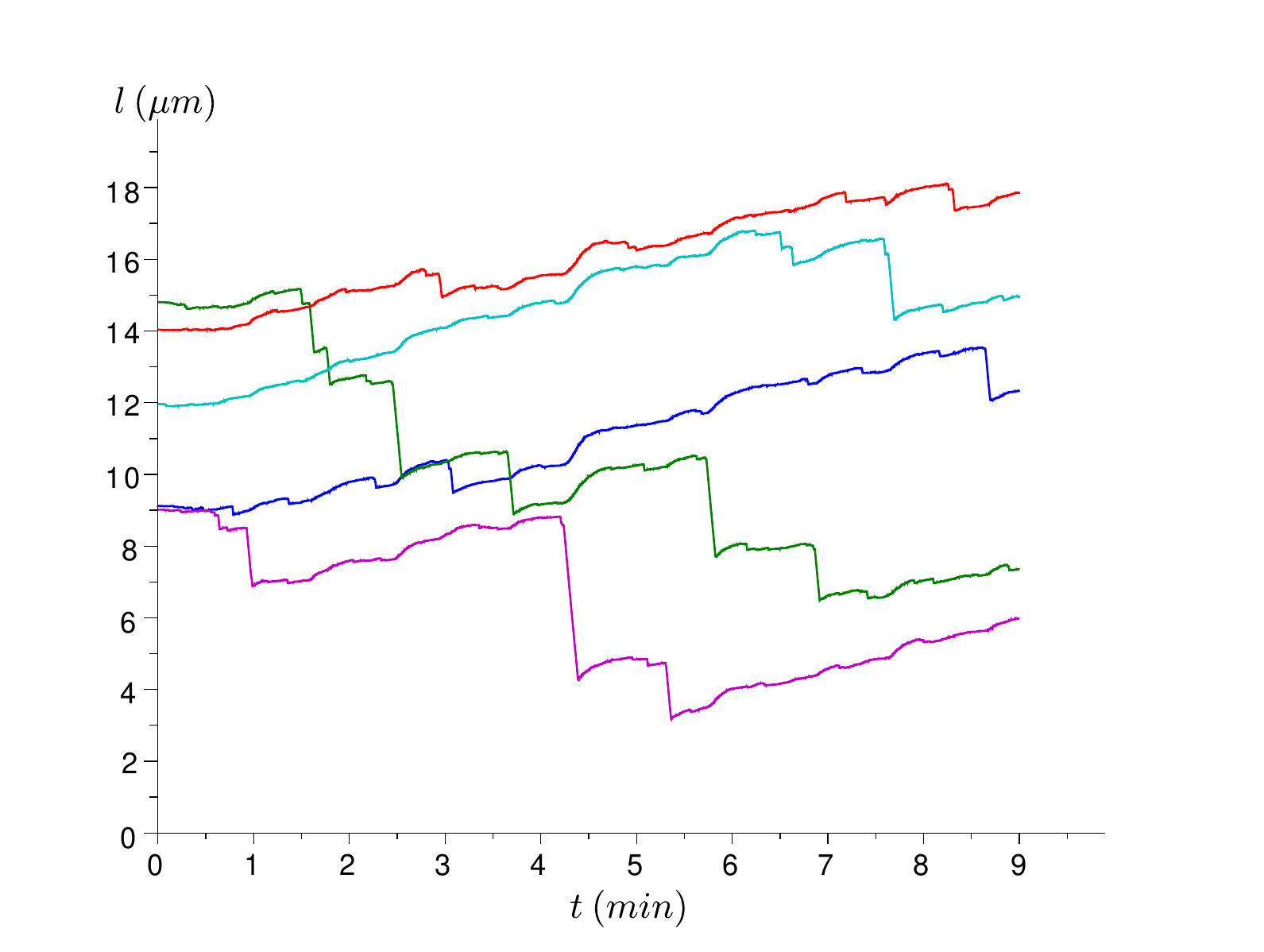}
\caption{Simulation of $m=5$ microtubules with a total of $\approx 10^5$
tubulin units in the presence a drug that completely inhibits GTP tubulin
hydrolysis. The relevant parameter values are $r=q=1$ and $s=0$. The total
amount of drug is $20000$ (left panel) respectively $40000$ (right panel). 
}\label{nonlocal_simulation}
\end{center}
\end{figure}

The nonlocal drug action mechanism where the bound drug inhibits hydrolysis of
GTP tubulin is much less effective in suppressing microtubule dynamic
instability. In Figure \ref{nonlocal_simulation} we show simulations of 5
microtubules built of approximately $10^5$ tubulin units in the presence of  a
drug that completely inhibits hydrolysis reactions (\ref{scalar}) and
(\ref{vectorial}), i.e.~$s=0$ without affecting the binding and loss reactions,
$r=q=1$. We observe that a much higher amount of drug, namely of the order of
tubulin units, is required to have any visible effect, while at the same time,
long periods of shortening still occur.

\begin{figure}[ht]
\begin{center}
\includegraphics[width=60mm]{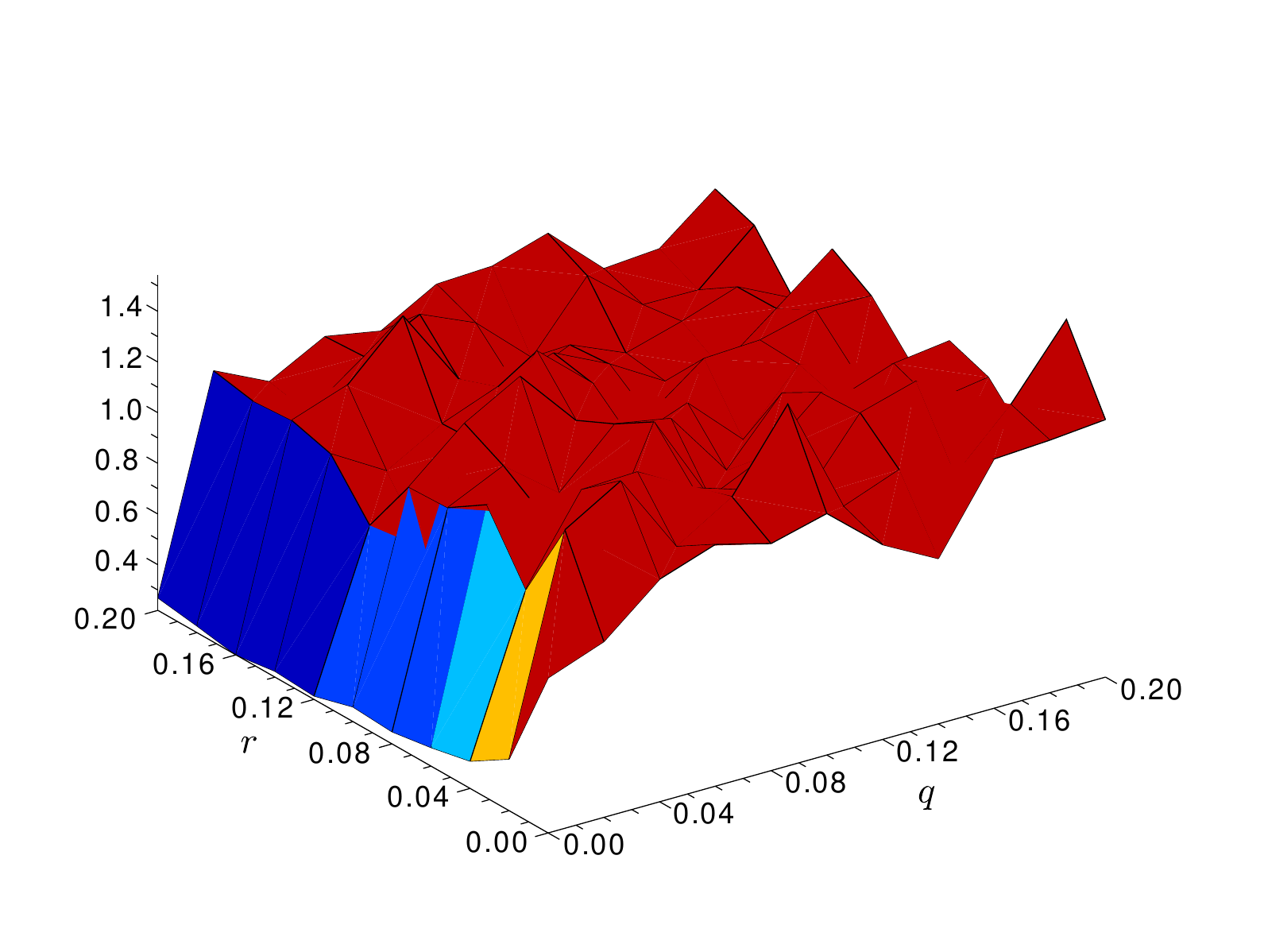}
\includegraphics[width=60mm]{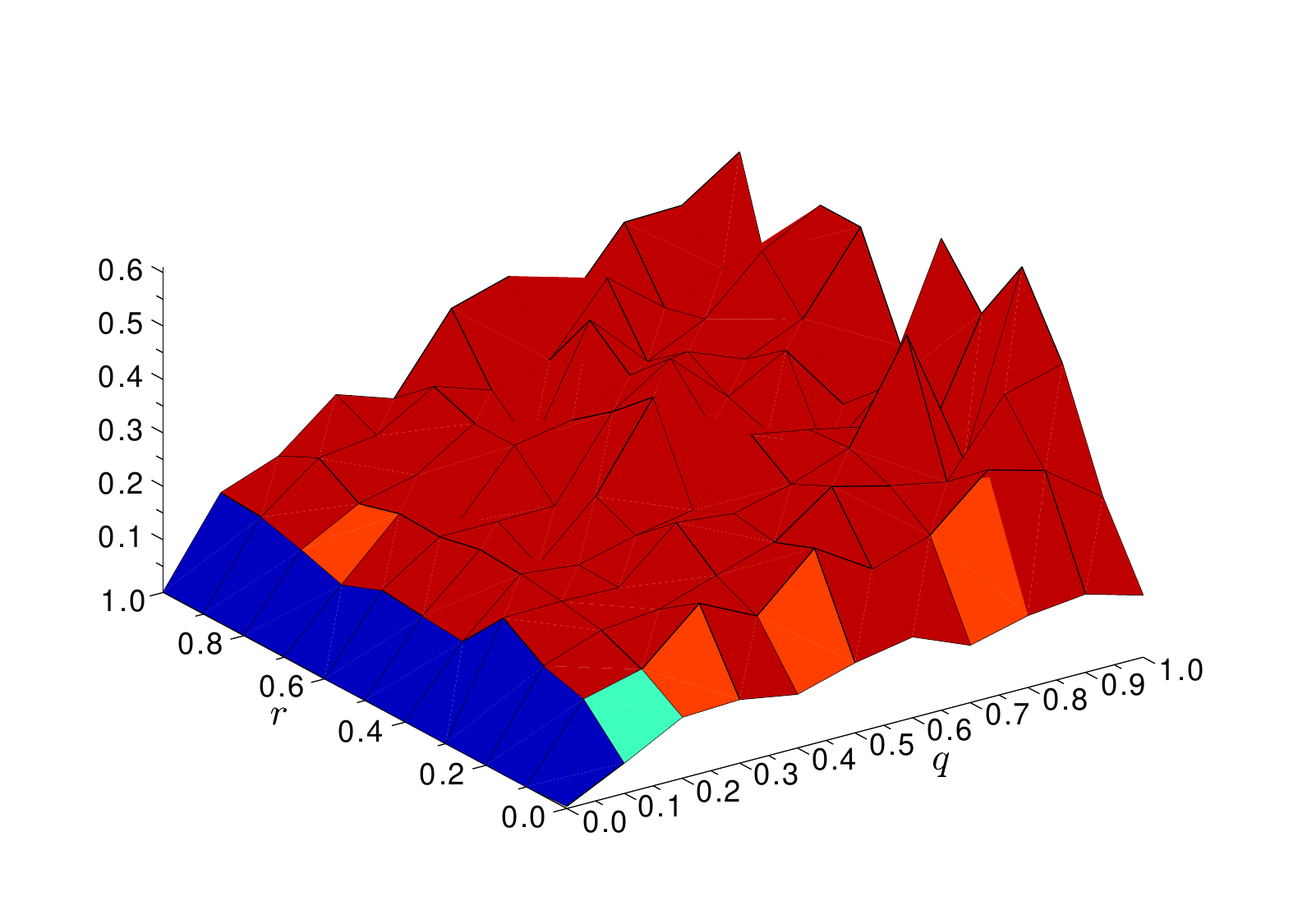}
\caption{The reduction of microtubule dynamic instability relative to the
untreated case as the drug effects vary. Shown are the relative peak heights of
the averaged absolute Fourier spectra of 20 microtubules at a concentration of
$100$ drug molecules for every microtubule. The drug molecules bind either at
any open site along the microtubule (left panel) or at the tip only (right
panel). The drug does not affect the hydrolysis of bound GTP
tubulin ($s=1$). }\label{reductionsurface}
\end{center}
\end{figure}

In order to better understand the influence of a drug acting at the microtubule
tip, we systematically varied the parameters $r$ and $q$, keeping all other
parameters and the drug concentration constant. We consider the reduction
of the peak height of the absolute Fourier spectra relative to the control
scenario. If the drug molecules are free to bind any of the open drug binding
sites, then a complete repression of the loss reactions (\ref{loss_GDP}) and
(\ref{loss_GTP}), i.e.~$q=0$ is required to efficiently suppress the dynamic
instability of microtubules, see Figure \ref{reductionsurface}, left panel. An
alternative is to allow drug molecules to bind only at the tip \cite{Lopus10b}.
In that case, the suppression effect persists for weaker drug effects
(Figure \ref{reductionsurface}, right panel). However, it is still more
important to suppress the loss reactions (small value of $q$) than to  suppress
the growth reaction ($r\approx 1$ is admissible). Notice that
these predictions of the model are drawn from the shapes of the surfaces in
Figure \ref{reductionsurface}, not from particular values. 

\section{Discussion}

Spatial and temporal regulation of the dynamic instability of microtubules is
essential to carry out several vital cellular functions. In cells, the
dynamicity of microtubules is regulated by a number of proteins such as MAPs
\cite{Lopus09}, G proteins \cite{Dave}, and the plus end tracking proteins,
including EB1 \cite{Lopus09}. Perturbations in the innate dynamicity of
microtubules induce cell cycle arrest and thereby inhibit cell proliferation.
Thus, compounds that target microtubules are potential anticancer drugs. Our
recent studies have found synthetic derivatives of maytansine such as DM1 (for
drug maytansinoid 1) that can be conjugated to tumor-specific antibodies as
potent suppressors of microtubule dynamics \cite{Oroudjev, Lopus10b}.
Antibody-DM1 Conjugates are under clinical evaluation, and they show promising
early results. Thus, the synthetic derivative of maytansine S-methyl-D-DM1 is a
good example to complement our modeling studies.

In this paper we have presented a detailed reaction scheme for microtubule
polymerization, GTP tubulin hydrolysis, catastrophic shrinking, recycling and
interaction with tubulin-binding drugs. Our \textit{in silico} simulations
show good agreement of Fourier spectra with experimental data of growing
microtubules under control conditions and treated with the maytansine derivative
S-methyl-D-DM1. Our model allows to accommodate a wide
variety of drug binding
mechanisms and interactions of the drug with the normal microtubule
polymerization and depolymerization processes. We find that drugs that act at
the microtubule tip by inhibiting addition of GTP tubulin and loss of GDP
tubulin are effective suppressors of microtubule dynamic
instability. This would also be the action mechanism of
S-methyl-D-DM1. Among these two actions, the inhibition of the loss of GDP
tubulin
is more important than the inhibition of the growth reaction. A localized
binding to the tip inhibits dynamic instability even more. On the other hand,
drugs whose action is to inhibit hydrolysis of bound GTP tubulin need to be
present at numbers comparable to the number of tubulin units to have a visible
effect on microtubule dynamic instability.

Our assumption in modeling the drug binding reaction (vii) has been
that either the drug molecule binds to any open binding site with equal
probability or that it binds only at an open site at the tip. The former is the
binding mode of a drug like paclitaxel that stabilizes a microtubule
along its entire length. Other drugs, such as vinblastine bind with high
affinity only at the microtubule plus end \cite{Jordan07}. In future work we
will implement a probability of drug binding that decreases with increasing
distance from the tip.

It is well known that some drugs have different effects at different
concentrations. For example, taxol (paclitaxel) increases microtubule
polymerization at high concentrations (50 taxol molecules per
100 tubulin molecules), while it reduces the rate of shortening at low
concentrations (1
taxol molecule per 100 tubulin molecules) \cite{Jordan07}. This suggests the
need for a ``nonlocal'' generalization of the perturbation of the binding and
shrinking processes, in contrast to the present choices in Equations
(\ref{drug_growth}) and (\ref{drug_shrink}).

The design of therapeutic drugs requires long and tedious searches among all
possible binding sites on the target. Identifying the most favorable
binding site (with the lowest binding energy), however, is needed to design or
discover a drug compound with the highest possible efficacy. Information on
drug-ligand binding affinities (for
tubulin and tubulin-binding drugs in particular) can be extracted by careful
comparison of simulation results and experimental data. Such estimations of
binding energies will be addressed in a future study. 

\ack Part of this research was carried out during visits at the University of
Wisconsin -- Milwaukee and the University of Alberta, Edmonton. We thank our
respective institutions for their warm hospitality. PH is partially supported by
NSF grant DMS-1016214. Financial support was partially supplied by grants 
from NSERC, Alberta's Advanced Education and Technology, the Allard Foundation
and the Alberta Cancer Foundation to JAT and from grant NIH CA57291 and a gift
from ImmunoGen Inc.~to MAJ. We thank Dr.~Ravi Chari (ImmunoGen Inc.,
Waltham, MA) for providing us with the drug \mbox{S-methyl-D-DM1} and Bruce
Fenske and Philip Winter (University of Alberta) for implementing the Gillespie
algorithm. We are much indebted to two anonymous readers whose comments greatly
helped to improve the paper.

\section*{References}
\bibliography{manuscript}

\begin{thebibliography}{10}

\bibitem{Jackson}
{Jackson, M. B.} and S.~A. Berkovitz.
\newblock Nucleation and the kinetics of microtubule assembly.
\newblock {\em Proc. Natl. Acad. Sci. USA}, {\bf 77}:7302--7305, 1980.

\bibitem{Mandelkow89}
{Mandelkow, E.}, E.~M. Mandelkow, H.~Hotani, B.~Hess, and S.~C. {M\"{u}ller}.
\newblock Spatial patterns from oscillating microtubules.
\newblock {\em Science}, {\bf 246}:1291--1293, 1989.

\bibitem{Walker88}
{Walker, R. A.}, E.~T. O'Brien, N.~K. Pryer, M.~E. Soboeiro, W.~A. Voter, H.~P.
  Erickson, and E.~D. Salmon.
\newblock Dynamic instability of individual microtubules analyzed by video
  light microscopy: rate constants and transition frequencies.
\newblock {\em J. Cell Biol.}, {\bf 107}:1437--1448, 1988.

\bibitem{BLT99}
{Bolterauer, H.}, {H.--J.} Limbach, and J.~A. Tuszy\'nski.
\newblock Microtubules: strange polymers inside the cell.
\newblock {\em Bioelectrochem. Bioenerg.}, {\bf 48}:285--295, 1999.

\bibitem{BLT99b}
{Bolterauer, H.}, {H.--J.} Limbach, and J.~A. Tuszy\'nski.
\newblock Models of assembly and disassembly of individual microtubules:
  stochastic and averaged equations.
\newblock {\em J. Biol. Phys.}, {\bf 25}:1--22, 1999.

\bibitem{Mitchison}
{Mitchison, T.} and M.~Kirschner.
\newblock Dynamic instability of microtubule growth.
\newblock {\em Nature}, {\bf 312}:237--242, 1984.

\bibitem{FHL94}
{Flyvbjerg, H.}, T.~E. Holy, and S.~Leibler.
\newblock Stochastic dynamics of microtubules: A model for caps and
  catastrophes.
\newblock {\em Phys. Rev. Lett.}, {\bf 73}:2372 --2375, 1994.

\bibitem{FHL96}
{Flyvbjerg, H.}, T.~E. Holy, and S.~Leibler.
\newblock Microtubule dynamics: caps, catastrophes, and coupled hydrolysis.
\newblock {\em Phys. Rev. E}, {\bf 54}:5538--5560, 1996.

\bibitem{JWF97}
{Jobs, E.}, D.~E. Wolf, and H.~Flyvbjerg.
\newblock Modeling microtubule oscillations.
\newblock {\em Phys. Rev. Lett.}, {\bf 79}:519--522, 1997.

\bibitem{Rezania_BPJ}
{Rezania, V.}, O.~Azarenko, M.~A. Jordan, H.~Bolterauer, R.~F. Ludue{\~n}a,
  J.~T. Huzil, and J.~A. Tuszy\'nski.
\newblock Microtubule assembly of isotypically purified tubulin and its
  mixtures.
\newblock {\em Biophys. J.}, {\bf 95}:1--16, 2008.

\bibitem{Lopus09}
{Lopus, M.}, M.~Yenjerla, and L.~Wilson.
\newblock {\em Microtubule Dynamics}, volume~{\bf 3} of {\em Wiley Encyclopedia
  of Chemical Biology}, pages 153--160.
\newblock 2009.

\bibitem{vanBuren2002}
{van Buren, V.}, D.~J. Odde, and L.~Cassimeris.
\newblock Estimates of lateral and longitudinal bond energies within the
  microtubule lattice.
\newblock {\em Proc. Natl. Acad. Sci. USA}, {\bf 99}:6035--6040, 2002.

\bibitem{vanBuren2005}
{van Buren, V.}, L.~Cassimeris, and D.~J. Odde.
\newblock Mechanochemical model of microtubule structure and self-assembly
  kinetics.
\newblock {\em Biophys. J.}, {\bf 89}:2911--2926, 2005.

\bibitem{PRE}
{Hinow, P.}, V.~Rezania, and J.~A. Tuszy\'nski.
\newblock A continuous model for microtubule dynamics with collapse, rescue and
  nucleation.
\newblock {\em Phys.~Rev.~E}, \textbf{80}:031904, 2009.
\newblock \href{http://arxiv.org/abs/0811.2245}{\texttt{arXiv:0811.2245}}.

\bibitem{Bayley2}
{Bayley, P. M.}, M.~J. Schilstra, and S.~R. Martin.
\newblock A lateral cap model of microtubule dynamic instability.
\newblock {\em FEBS Letters}, {\bf 259}:181--184, 1989.

\bibitem{Antal}
{Antal, T.}, P.~L. Krapivsky, S.~Redner, M.~Mailman, and B.~Chakraborty.
\newblock Dynamics of an idealized model of microtubule growth and catastrophe.
\newblock {\em Phys. Rev. E}, {\bf 76}:041907, 2007.
\newblock
  \href{http://arxiv.org/abs/q-bio/0703001}{\texttt{arXiv:q-bio/0703001}}.

\bibitem{Gregoretti}
{Gregoretti, I. V.}, G.~Margolin, M.~S. Alber, and H.~V. Goodson.
\newblock Insights into cytoskeletal behavior from computational modeling of
  dynamic microtubules in a cell--like environment.
\newblock {\em J. Cell Sci.}, {\bf 119}:4781--4788, 2006.
\newblock
  \href{http://arxiv.org/abs/q-bio/0604023}{\texttt{arXiv:q-bio/0604023}}.

\bibitem{Matzavinos}
{Matzavinos, A.} and H.~G. Othmer.
\newblock A stochastic analysis of actin polymerization in the presence of
  twinfilin and gelsolin.
\newblock {\em J. Theor. Biol.}, {\bf 249}:723--736, 2007.

\bibitem{Brun}
{Brun, L.}, B.~Rupp, J.~Ward, and F.~Nedelec.
\newblock A theory of microtubule catastrophes and their regulation.
\newblock {\em Proc. Natl. Acad. Sci. USA}, {\bf 106}:21173--21178, 2009.

\bibitem{Mishra}
{Mishra, P. K.}, A.~Kunwar, S.~Mukherji, and D.~Chowdhury.
\newblock Dynamic instability of microtubules: {Effect} of
  catastrophe-suppressing drugs.
\newblock {\em Phys. Rev. E}, {\bf 72}:051914, 2005.
\newblock \href{http://arxiv.org/abs/cond-mat/0310546}{\texttt{
  arXiv:cond-mat/0310546}}.

\bibitem{Gillespie}
{Gillespie, D. T.}
\newblock Exact stochastic simulation of coupled chemical reactions.
\newblock {\em J. Phys. Chem.}, {\bf 81}:2340--2361, 1977.

\bibitem{Jordan07}
{Jordan, M. A.} and K.~Kamath.
\newblock How do microtubule-targeted drugs work? {An} overview.
\newblock {\em Current Cancer Drug Targets}, {\bf 7}:730--742, 2007.

\bibitem{Bennett}
{Bennett, M. J.}, J.~K. Chik, G.~W. Slysz, T.~Luchko, J.~A. Tuszy\'nski, D.~L.
  Sackett, and D.~C. Schriemer.
\newblock Structural mass spectrometry of the $\alpha\beta$-tubulin dimer
  supports a revised model of microtubule assembly.
\newblock {\em Biochemistry}, {\bf 48}:4858--4870, 2009.

\bibitem{Correia08}
{Correia, J. J.} and S.~Lobert.
\newblock Molecular mechanisms of microtubule acting cancer drugs.
\newblock In T.~Fojo, editor, {\em Microtubules in Health and Disease}, pages
  21--46. 2008.

\bibitem{Oroudjev}
{Oroudjev, E.}, M.~Lopus, L.~Wilson, C.~Audette, C.~Provenzano, H.~Erickson,
  Y.~Kovtun, R.~Chari, and M.~A. Jordan.
\newblock Maytansinoid-antibody conjugates induce mitotic arrest by suppressing
  microtubule dynamic instability.
\newblock {\em Mol. Cancer Therapeutics}, {\bf 9}:2700--2713, 2010.
\newblock doi:10.1158/1535-7163.MCT-10-0645.

\bibitem{Lopus11}
{Lopus, M.}
\newblock {Antibody-DM1 conjugates as cancer therapeutics}.
\newblock {\em Cancer Lett.}, {\bf 307}:113--118, 2011.

\bibitem{Mozziconacci08}
{Mozziconacci, J.}, L.~Sandblad, M.~Wachsmuth, D.~Brunner, and E.~Karsenti.
\newblock Tubulin dimers oligomerize before their incorporation into
  microtubules.
\newblock {\em PLoS ONE}, {\bf 3}:e3821, 2008.

\bibitem{Howard09}
{Howard, J.} and A.~A. Hyman.
\newblock Growth, fluctuation and switching at microtubule plus ends.
\newblock {\em Nat. Rev. Mol. Cell Biol.}, {\bf 10}:569--574, 2009.

\bibitem{Dimitrov}
{Dimitrov, A.}, M.~Quesnoit, S.~Moutel, I.~Cantaloube, C.~Po{\"u}s, and
  F.~Perez.
\newblock Detection of {GTP-Tubulin} conformation in vivo reveals a role for
  {GTP} remnants in microtubule rescues.
\newblock {\em Science}, {\bf 322}:1353--1356, 2008.

\bibitem{Lin1981}
{Lin, C. M.} and E.~Hamel.
\newblock Effects of inhibitors of tubulin polymerization on {GTP} hydrolysis.
\newblock {\em J. Biol. Chem.}, {\bf 256}:9242--9245, 1981.

\bibitem{Widdison}
{Widdison, W. C.}, S.~D. Wilhelm, E.~E. Cavanagh, K.~R. Whiteman, B.~A. Leece,
  Y.~Kovtun, V.~S. Goldmacher, H.~Xie, R.~M. Steeves, R.~J. Lutz, R.~Zhao,
  L.~Wang, W.~A. Bl{\"a}ttler, and R.~V.~J. Chari.
\newblock Semisynthetic maytansine analogues for the targeted treatment of
  cancer.
\newblock {\em J. Med. Chem.}, {\bf 49}:4392--4408, 2006.

\bibitem{Lopus10a}
{Yenjerla, M.}, M.~Lopus, and L.~Wilson.
\newblock {\em Analysis of Dynamic Instability of Steady-State Microtubules In
  Vitro by Video-Enhanced Differential Interference Contrast Microscopy},
  volume~{\bf 95} of {\em Methods in Cell Biology}, chapter~11, pages 189--206.
\newblock 2010.

\bibitem{Lopus10b}
{Lopus, M.}, E.~Oroudjev, L.~Wilson, S.~Wilhelm, W.~Widdison, R.~Chari, and
  M.~A. Jordan.
\newblock Maytansine and cellular metabolites of antibody-maytansinoid
  conjugates strongly suppress microtubule dynamics by binding to microtubules.
\newblock {\em Mol. Cancer Therapeutics}, {\bf 9}:2689--2699, 2010.
\newblock doi:10.1158/1535-7163.MCT-10-0644.

\bibitem{Yenjerla}
{Yenjerla, M.}, N.~E. La{P}ointe, M.~Lopus, C.~Cox, M.~A. Jordan, S.~C.
  Feinstein, and L.~Wilson.
\newblock The neuroprotective peptide {NAP} does not directly affect
  polymerization or dynamics of reconstituted neural microtubules.
\newblock {\em J. Alzheimer Dis.}, {\bf 19}:1377--1386, 2010.

\bibitem{Odde1996}
{Odde, D. J.}, H.~M. Buettner, and L.~Cassimeris.
\newblock Spectral analysis of microtubule assembly dynamics.
\newblock {\em AIChE Journal}, {\bf 42}:1434--1442, 1996.

\bibitem{Dave}
{Dav{\'e}, R. H.}, W.~Saengsawang, S.~Dav{\'e}, M.~Lopus, L.~Wilson, and M.~M.
  Rasenick.
\newblock A molecular and structural mechanism for {G}-protein mediated
  microtubule destabilization.
\newblock {\em J.~Biol.~Chem}, {\bf 286}:4319--4328, 2011.

\bibitem{Carlier}
{Carlier, M. F.}, R.~Melki, D.~Pantaloni, T.~L. Hill, and Y.~Chen.
\newblock Synchronous oscillations in microtubule polymerization.
\newblock {\em Proc. Natl. Acad. Sci. USA}, {\bf 84}:5257, 1987.

\end{thebibliography}

\section{Tables and Figures}
\begin{table}[th]
\begin{center}
\begin{tabular}{|c|c|l|c|}\hline\hline
parameter & value  & remark & reference \\
\hline
$\lambda$ &  $0.4 \,(\ell s)^{-1}$ & addition rate of GTP tubulin to GTP tip
&  \cite{Antal} \\ 
$p$ &  $0.05$ & reduction of GTP addition to GDP tip & \cite{Antal} \\
$\mu_{GDP}$ &  $800  \,s^{-1}$ & loss of GDP tubulin from tip &
\cite{Walker88} \\
$\mu_{GTP}$ &  $1.5  \,s^{-1}$ & loss of GTP tubulin from tip & \cite{FHL96,
Howard09}\\
$\delta_{sc}$ &  $1.2  \,(\ell s)^{-1}$  & rate of scalar hydrolysis &
\cite{FHL96} \\
$\delta_{vec}$ &  $1.2  \,(\ell s)^{-1}$ & rate of vectorial hydrolysis &
\cite{FHL96} \\
$\kappa$ &  $0.1\, s^{-1}$ & rate of GDP tubulin recycling & \cite{Carlier} \\
$L$ &  $6$ & average GTP tubulin addition size & \cite{Mozziconacci08} \\
\hline
$\rho$ &  $1.0  s^{-1}$ & rate of drug binding &   \\
$\sigma$ &  $0.1  \, s^{-1}$ & rate of drug release &   \\
\hline
\end{tabular}
\caption{Baseline values of parameters used in the stochastic simulations. Here
$\ell = 0.6\,nm$ is the gain in length by addition of a single tubulin unit. The
references provide further discussion and sometimes comparable values. 
}\label{Tab1}
\end{center}
\end{table}


\end{document}